\documentclass[aps,pra,twocolumn,showpacs,floatfix,amsmath]{revtex4}

\usepackage{graphicx}
\usepackage{hyperref}
\usepackage{bm}

\newcommand{\bNi}{\mathbf{N}_{i}}
\newcommand{\bfN}{\mathbf{N}}
\newcommand{\meanP}{\overline{P}}
\newcommand{\specdenN}{S_{N}(f)}
\newcommand{\meann}{\overline{n}}

\newcommand{\sgn}{\mathrm{sgn}}
\newcommand{\bfsig}{\mbox{\boldmath $\scriptstyle\sigma$}}
\newcommand{\bigGama}{\Gamma_{\mbox{\boldmath $\scriptstyle\sigma ,
                       \scriptstyle\sigma^{\prime}$}}}
\newcommand{\decofac}{D_{\mbox{\boldmath $\scriptstyle\sigma ,
                       \scriptstyle\sigma^{\prime}$}}}
\newcommand{\bfxh}{\hat{\mathbf{x}}}
\newcommand{\bfyh}{\hat{\mathbf{y}}}

\begin{document}

\title{Simulation of Quantum Adiabatic Search in the Presence of Noise} 

\date{\today}

\author{Frank Gaitan}
\affiliation{Department of Physics; Southern Illinois University;
Carbondale, IL 62901-4401}

\begin{abstract}
Results are presented of a large-scale simulation of the quantum adiabatic 
search (QuAdS) algorithm in the presence of noise. 
The algorithm is applied to the $NP$--Complete problem N--Bit Exact Cover 3 
(EC3). The noise is assumed to Zeeman-couple to the qubits and its
effects on the algorithm's performance is studied for various levels of noise 
power, and for 4 different types of noise polarization. We examine the 
scaling relation between the number of bits $N$ (EC3 problem-size) and the 
algorithm's noise-averaged median run-time $\langle {T}(N)\rangle$. Clear 
evidence is found of the algorithm's sensitivity to noise. Two fits to the
simulation results were done: (1) power-law scaling $\langle T(N) \rangle
= aN^{b}$; and (2) exponential scaling $\langle T(N) \rangle = a\left[
\exp (bN)-1\right]$. Both types of scaling relations provided excellent fits, 
although the scaling parameters $a$ and $b$ varied with noise power, and with 
the type of noise polarization. The sensitivity of the scaling exponent $b$ to 
noise polarization allows a relative assessment of which noise polarizations 
are most problematic for quantum adiabatic search. We demonstrate how the 
noise leads to decoherence in QuAdS, and estimate the amount of decoherence 
present in our simulations. An upper bound is also derived for the 
noise-averaged QuAdS success probability in the limit of weak noise that is
appropriate for our simulations.
\end{abstract}


\maketitle

\section{\label{sec1}Introduction}

In computational complexity theory \cite{pap}, computational problems are
classified according to the resources needed to obtain a solution. Often
such a problem is reformulated as a decision problem whose
solution is a ``yes'' or ``no'' answer. An algorithm that solves all
instances of a decision problem is said to be a polynomial-time algorithm
if the time needed to find a solution to an arbitrary instance is 
$\mathcal{O}(n^{k})$, where $n$ is 
the size of the instance and $k$ is a fixed positive integer. The 
computational complexity class $P$ is composed of decision problems for
which polynomial-time algorithms exist. Another important computational
complexity class is the class $NP$ which is composed of decision problems
for which, for each such problem, a polynomial-time algorithm exists to verify 
the ``yes'' output of a candidate solution. The question of whether the 
classes $P$ and $NP$ are equal is one of the biggest open problems in 
theoretical computer science. It is widely conjectured that $P\neq NP$. 
A decision problem $p$ is said to be polynomially transformable to a decision 
problem $q$ if: (1) there exists a function $f$ that maps every instance $x$ 
of $p$ into an instance $f(x)$ of $q$ for which the answer to $x$ is ``yes'' 
if and only if the answer to $f(x)$ is ``yes''; and (2) a polynomial-time 
algorithm exists to compute $f(x)$ for every $x$ in $p$. A problem is said to 
be $NP$--Complete if it belongs to $NP$ and every problem in $NP$ is
polynomially transformable to it. Thus, if a polynomial-time algorithm is
found for an $NP$--Complete problem, it follows that $P=NP$. In light of
Shor's polynomial-time quantum factoring algorithm \cite{sho}, Bennett et.\
al.\ \cite{ben} asked whether a polynomial-time quantum algorithm might
exist to solve an $NP$--Complete problem. In other words, they asked, ``Can a 
quantum computer solve an $NP$--Complete problem in polynomial-time?'' Should 
the answer to this question prove to be ``yes'', it is widely recognized that 
this would have profound consequences for theoretical computer 
science \cite{n+c}.

In 2001, Farhi et.\ al.\ \cite{fa1} examined the question of whether a
quantum computer might be able to solve an $NP$--Complete problem in
polynomial-time. They applied the quantum adiabatic search (QuAdS) algorithm
introduced in Ref.~\cite{fa2} to find solutions of randomly generated hard
instances of the $NP$--Complete problem N-Bit Exact Cover~3 which they  
believe to be classically intractable for sufficiently large inputs. 
The algorithm was simulated numerically on an existing (classical) computer. 
Because a quantum computer cannot be simulated efficiently on a classical 
computer, the simulations were restricted to $7\leq N \leq 20$. They found 
that the median run-time $\overline{T}(N)$ for QuAdS to succeed on this
class of instances could be fit with a quadratic scaling relation in
the number of bits $N$ (problem-size): $\overline{T}(N)\sim N^{2}$. It was 
pointed out that should classical algorithms truly require exponential time 
on this class of instances, and should the quadratic scaling behavior of 
QuAdS persist to large $N$, then QUAdS could outperform classical algorithms 
on randomly generated hard instances, though not necessarily on worst case 
instances. Their work has drawn a great deal 
of interest and suggests two possible directions for further research: 
(1) examining the large $N$ scaling behavior of QuAdS; and (2) examining the 
robustness of QuAdS to noise. Recent progress on (1) appears in the work of 
Roland and Cerf \cite{r+c}; Mitchell et.\ al.\ \cite{mit}; and Latorre
and Or\'{u}s \cite{LO1,LO2}. In this paper we will focus on the question of 
robustness of QuAdS to noise.

The analysis of Ref.~\cite{fa1} assumed that the quantum computer was perfectly
isolated from noise. Childs et.\ al.\ \cite{chi} were the first to consider 
the robustness of QuAdS to noise. They presented numerical simulation results 
which suggested QuAdS might have an inherent robustness to two types of 
noise-induced errors: (1) transitions out of the instantaneous ground state; 
and (2) unitary control errors. Their simulation did not, however, attempt 
to match the scale of the simulation presented in Ref.~\cite{fa1}. Recently, 
Roland and Cerf carried out an analytical study of QuAdS in the presence of 
noise using random matrix theory \cite{rc2}. They showed that the error 
probability of QuAdS would not increase with problem-size $N$ if the noise 
varies either very slowly or very rapidly with respect to the natural 
time-scale of the quantum computation
$\hbar /\overline{E}$, where $\overline{E}$ is the energy scale for the
eigenvalues of the instantaneous Hamiltonian $H(t)$. Their analysis assumed
weak noise which justified a perturbative analysis of their model.

As stated above, in this paper we will also examine the robustness of QuAdS 
to noise. Our work
complements the analysis of Ref.~\cite{rc2} in that we consider noise that 
varies on a time-scale comparable to $\hbar /\overline{E}$. We extend the
simulation protocol of Farhi et.\ al.\ \cite{fa1} to include noise, and
determine the noise-averaged median runtime $\langle{T}(N)\rangle$ for QuAdS 
to succeed on randomly generated instances of N-Bit Exact Cover 3
which have a unique solution. We find clear evidence of the sensitivity 
of QuAdS to noise. Two fits to the simulation results were carried out:
(1) power-law scaling $\langle T(N) \rangle = aN^{b}$; and (2) exponential
scaling $\langle T(N) \rangle = a\left[\exp (bN)-1\right]$. Both types of
scaling relations provided excellent fits, although the scaling parameters
$a$ and $b$ varied with noise power, and with the noise polarization (defined
in Section~\ref{sec3a}). Although we did encounter examples of noise 
realizations that reduce the runtime for QuAdS to succeed, at the largest 
noise power levels that we examined, the predominant effect of 
noise was to slowdown QuAdS. The sensitivity of QuAdS to noise 
polarization allows, for the first time, a determination of which noise
polarizations are most problematic for QuAdS.

The structure of this paper is as follows. In section~\ref{sec2} we briefly
describe the $NP$--Complete problem N-Bit Exact Cover 3, and review how 
QuAdS can be used to find a solution to an instance of this problem. In 
section~\ref{sec3} we describe our noise model and explain our extension of
the simulation protocol of Farhi et.\ al.\ to include noise. In 
section~\ref{sec4} we present the results of our simulation, and discuss their 
significance in section~\ref{sec5}. Section~\ref{sec5} also demonstrates
how noise-induced dephasing leads to decoherence in QuAdS, estimates the
amount of decoherence present in our simulations, and derives an upper
bound for the noise-averaged QuAdS success probability in the weak noise
limit that is appropriate for our simulations. Finally, in section~\ref{sec6}, 
we provide a summary of our work.

\section{\label{sec2} Exact Cover 3 and Quantum Adiabatic Search}

In this section we: (i) present the $NP$-Complete problem Exact Cover~3
(section~\ref{sec2a}); (ii) describe the quantum adiabatic search algorithm
and how it is used to solve instances of Exact Cover~3 (section~\ref{sec2b});
and (iii) describe the simulation protocol of Ref.~\cite{fa1} 
(section~\ref{sec2c}).

\subsection{\label{sec2a} Exact Cover 3}

Consider a collection of $N$ binary variables (bits) $z_{1},\cdots\: ,z_{N}$
each of which can take on the value $0$ or $1$. The state of this $N$-bit
system is specified by assigning values to each bit in the string 
$z=z_{1}\cdots z_{N}$. A total of $2^{N}$ states $z$ are possible.

A specific instance of Exact Cover~3 (EC3) is composed of $L$ clauses, each
of which imposes a constraint on the values of three of the $N$ bits. The 
number of clauses $L$ will generally vary from one EC3-instance to another. 
If the $i$-th clause involves the bits ($a(i)$, $b(i)$, $c(i)$), then a 
particular bit-string $z=z_{1}\cdots z_{N}$ satisfies the clause if $z_{a(i)}+
z_{b(i)}+z_{c(i)} = 1$. Otherwise, one says that $z$ does not satisfied the 
clause. A bit-string $z$ is said to solve an instance of EC3 if it satisfies 
all the clauses that make up the instance. If one is given an instance of EC3
and a candidate solution string $z^{\ast}$, one can check whether $z^{\ast}$ 
solves the instance in polynomial-time since checking each of the $L$ clauses 
simply requires adding three integers, and this can be done in 
polynomial-time. Thus EC3 belongs to $NP$. It can be shown that all problems 
in $NP$ are polynomially transformable to EC3 \cite{pa2} so that EC3 is also 
$NP$-Complete.

\subsection{\label{sec2b} QuAdS and EC3}

By combining the dynamics of the quantum adiabatic theorem with a clever 
choice of Hamiltonian $H(t)$, quantum adiabatic search (QuAdS) causes
the state of a quantum computer (QC) to home-in on a solution of an instance
of an $NP$-Complete problem \cite{fa1}. We assume the $NP$-Complete problem 
is EC3. The quantum adiabatic theorem \cite{mes} assures us that if the 
initial state of a quantum system is the groundstate $|E_{g}(0)\rangle$ of 
the initial Hamiltonian $H(0)$, and $H(t)$ varies sufficiently slowly, 
then the quantum state at time $T$ can be brought arbitrarily close to 
the groundstate of $H(T)$: $|\psi (T)\rangle = |E_{g}(T)\rangle + 
|\delta\psi\rangle$. The probability not to be in the groundstate, 
$\epsilon (T) = |\langle\delta\psi|\delta\psi\rangle |^{2}$, vanishes as 
$T\rightarrow\infty$. For a given instance of EC3, QuAdS evolves the state 
$|\psi (t)\rangle$ over a time interval $0\leq t \leq T$, and the final
Hamiltonian $H(T)=H_{p}$ is constructed so that its groundstate encodes a
solution of the given instance. The initial Hamiltonian $H(0)=H_{i}$ is chosen 
so that its groundstate is non-degenerate and can be easily constructed,
and $H(t)$ smoothly morphs $H_{i}\rightarrow H_{p}$:
\begin{equation}
H(t) = \left(\, 1 -\frac{t}{T}\,\right)\, H_{i} + \left(\frac{t}{T}\right)\, 
        H_{p} \hspace{0.1in} .
\label{timeHam}
\end{equation}
We now briefly describe how $H_{i}$ and $H_{p}$ are constructed for a given
instance of EC3.

Suppose that we are given an instance of $N$-bit EC3 composed of $L$ clauses
$C_{i} = (\, a(i),\: b(i),\: c(i)\, )$ with $i=1,\cdots , L$, and that the
bit-string $s = s_{1}\cdots s_{N}$ is a solution of this instance. Our QC is
assumed to contain $N$ qubits and the computational basis states 
$|z_{1}\cdots z_{N}\rangle$ are chosen to be eigenstates of $\sigma_{z}^{1}
\otimes\cdots\otimes\sigma_{z}^{N}$. Although the QuAdS algorithm can be 
formulated more abstractly, it proves convenient to adopt the language of
NMR. $H_{i}$, then, describes the Zeeman coupling of the qubits to an 
external magnetic field which points along the x-direction. The strength of 
the magnetic field at the site of the $i$-th qubit is equal to the number of 
clauses in the EC3 instance that contain bit $i$. The non-degenerate 
groundstate of $H_{i}$ thus has all qubit spins aligned along 
$\hat{\mathbf{x}}$ so that $|E_{g}(0)\rangle$ is simply the uniform 
superposition of all $2^{N}$ computational basis states
$|z_{1}\cdots z_{N}\rangle$. The final Hamiltonian $H_{p}$ is constructed to
be diagonal in the computational basis $|z_{1}\cdots z_{N}\rangle$:
\begin{equation}
H_{p}\, |z_{1}\cdots z_{N}\rangle = h(z_{1},\cdots ,z_{N})\, 
                                       |z_{1}\cdots z_{N}
                                               \rangle \hspace{0.1in} .
\end{equation}
The eigenvalue $h(z_{1},\cdots , z_{N})$ is the sum of energy functions
$h_{C_{i}}(z_{a(i)},z_{b(i)},z_{c(i)})$, 
\begin{equation}
h(z_{1},\cdots ,z_{N}) = \sum_{i=1}^{L}\, h_{C_{i}}(z_{a(i)},z_{b(i)},
                            z_{c(i)}) \hspace{0.1in} ,
\end{equation}
where $C_{i}=(a(i),b(i),c(i))$
is the $i$-th clause in the given EC3 instance and,
\begin{displaymath}
h_{C_{i}}(z_{a(i)},z_{b(i)},z_{c(i)}) = 
                    \left\{ \begin{array}{cc}
                       0 & \hspace{0.1in}
                             \mbox{if $z_{1}\cdots z_{N}$ satisfes $C_{i}$} \\
                       1 & \mbox{otherwise  .}
                    \end{array} \right.   
\end{displaymath}
Thus $h(z_{1},\cdots ,z_{N})$ indicates how many clauses 
are violated by the string $z = z_{1}\cdots z_{N}$. All strings 
$s=s_{1}\cdots s_{N}$ that solve the given instance correspond to 
computational basis states $|s_{1}\cdots s_{N}\rangle$ that have zero energy,
and which together span the groundstate eigenspace of $H_{p}$. EC3 instances 
which have a unique solution are referred to as unique satisfying assignment 
(USA) instances, and are believed to be the most difficult for QuAdS.
In essence, it's harder to find a needle in a haystack that contains only 
one needle, than it is to find one in a haystack that contains many needles.
For further discussion of instances with multiple satisfying assignments,
see Ref.~\cite{non}.

Thus, if our QC is initially prepared in the groundstate of $H_{i}$, and $T$ 
is chosen to be sufficiently large that $H(t)$ evolves adiabatically, then the 
final state $|\psi (T)\rangle$ will lie in the groundstate eigenspace of 
$H_{p}$ with probability $1 - \epsilon (T)$, where $\epsilon (T) = 
|\langle\delta\psi |\delta\psi\rangle |^{2} 
\ll 1$. For a USA instance with solution string $s=s_{1}\cdots s_{N}$, 
measuring $|\psi (T)\rangle$ in the computational basis will give the solution 
string $s$ with probability $1 - \epsilon (T)$. Let the string $z_{1}\cdots
z_{N}$ be the actual measurement result. It can be quickly tested to determine 
whether it solves the USA instance. If ``yes'', the algorithm has succeeded, 
and $z=s$. If ``no'', then the quantum adiabatic search procedure is repeated 
until the measurement result $z = s$. The algorithm's failure probability 
$\epsilon (T)$ will be much less than $1$ if \cite{mes}:
\begin{equation}
T \gg \frac{\varepsilon}{\Delta^{2}} \hspace{0.1in} ,
\end{equation}
where,
\begin{eqnarray}
\varepsilon  & = & \max_{t\:\epsilon\, [0, T]}\: \left|\,
                \langle E_{1}(t)|\, T\,\frac{dH}{dt}|E_{g}(t)\rangle\, 
                 \vspace{0.5in}  \right| \hspace{0.1in} ; \\
\Delta  & = & \min_{t\:\epsilon\: [0, T]}\: |E_{1}(t)-E_{g}(t)| 
                   \hspace{0.1in} ; \label{mingap} 
\end{eqnarray}
and
$|E_{g}(t)\rangle$ ($|E_{1}(t)\rangle$) is the groundstate (first excited
state) of $H(t)$.

\subsection{\label{sec2c} Noiseless Simulation Protocol}

Farhi and co-workers \cite{fa1,fa2} studied the performance of QuAdS by
numerically integrating the Schrodinger equation using $H(t)$ 
(eqn.~(\ref{timeHam})) to drive the dynamics. Because the Hilbert space for
$N$-qubits has dimension $2^{N}$, numerical simulation quickly becomes 
impractical since the number of wavefunction components that must be 
tracked by the numerical integration grows exponentially with $N$. This
practical difficulty caused the simulations done in Refs.~\cite{fa1,fa2} to 
be restricted to the range $7\leq N \leq 20$. Only randomly generated USA 
instances were simulated as these authors believed that this set would 
provide hard cases for QuAdS, though not necessarily the worst cases.

An $N$-bit USA instance of EC3 is generated by the following procedure. 
The first clause is generated by picking 3 distinct integers in the range
$[ 1,\: N ]$ at random (uniform deviate). A count of the number of bit-strings 
of length $N$ that satisfy the clause is then done. Distinct clauses continue 
to be generated in this manner, and with each new clause generated, a new count of the number of bit-strings that satisfy all clauses generated to that point 
is done. Eventually, the point is reached where enough clauses have been 
generated so that one of two situations occurs. (1)~Only one bit-string $s$ of 
length $N$ remains that satisfies all the clauses generated. The collection of 
clauses then corresponds to an EC3-instance which has a unique solution $s$, 
and the procedure has thus produced a USA instance of EC3. (2)~Addition of 
the most recently generated clause causes the set of all clauses produced to
have no satisfying assignment (i.~e.\ an EC3-instance has been produced with 
no solution). In this case, the instance is discarded and the above procedure 
is repeated until a USA instance is generated. The USA instance 
produced is then used to construct $H_{i}$ and $H_{p}$, and from them, $H(t)$.
The USA instances generated by this procedure were found to contain 
approximately as many clauses $L$ as number of bits $N$: $L\sim N$. 

In the adiabatic limit, $T\rightarrow\infty$, the QuAdS success probability 
$P_{s}\rightarrow 1$. Numerical 
simulations must necessarily work with finite $T$ so that a protocol is
needed to determine how to pick $T$. Farhi et.~al.\ carried out a hunting
procedure which searches for a $T$ that causes the simulation to produce
a success probability $P_{s}$ in the range $[ 0.12,\: 0.13 ]$. This 
corresponds to a success probability in the vicinity of $P_{s} = 0.125= 
1/2^{3}$. As noted in Ref.~\cite{fa1}, this value of $P_{s}$ is somewhat
arbitrary, although for the values of $N$ used in the simulations, it is
much larger than $1/2^{N}$ which is the probability of the solution string
$s$ in the initial state. For each value of $N$, 75 USA instances are
generated and QuAdS is used to find the solution to each instance. This
generates runtimes $T_{i}(N)$ ($i= 1, \cdots , 75$) from which the median
runtime $\overline{T}(N)$ is determined. Farhi et.\ al.\ \cite{fa1,fa2} 
found that their simulation results for $\overline{T}(N)$ could be fit with a 
quadratic scaling relation for the range of $N$-values considered:
$\overline{T}(N)\sim N^{2}$. 

\begin{figure}[!]
\includegraphics[scale=0.325]{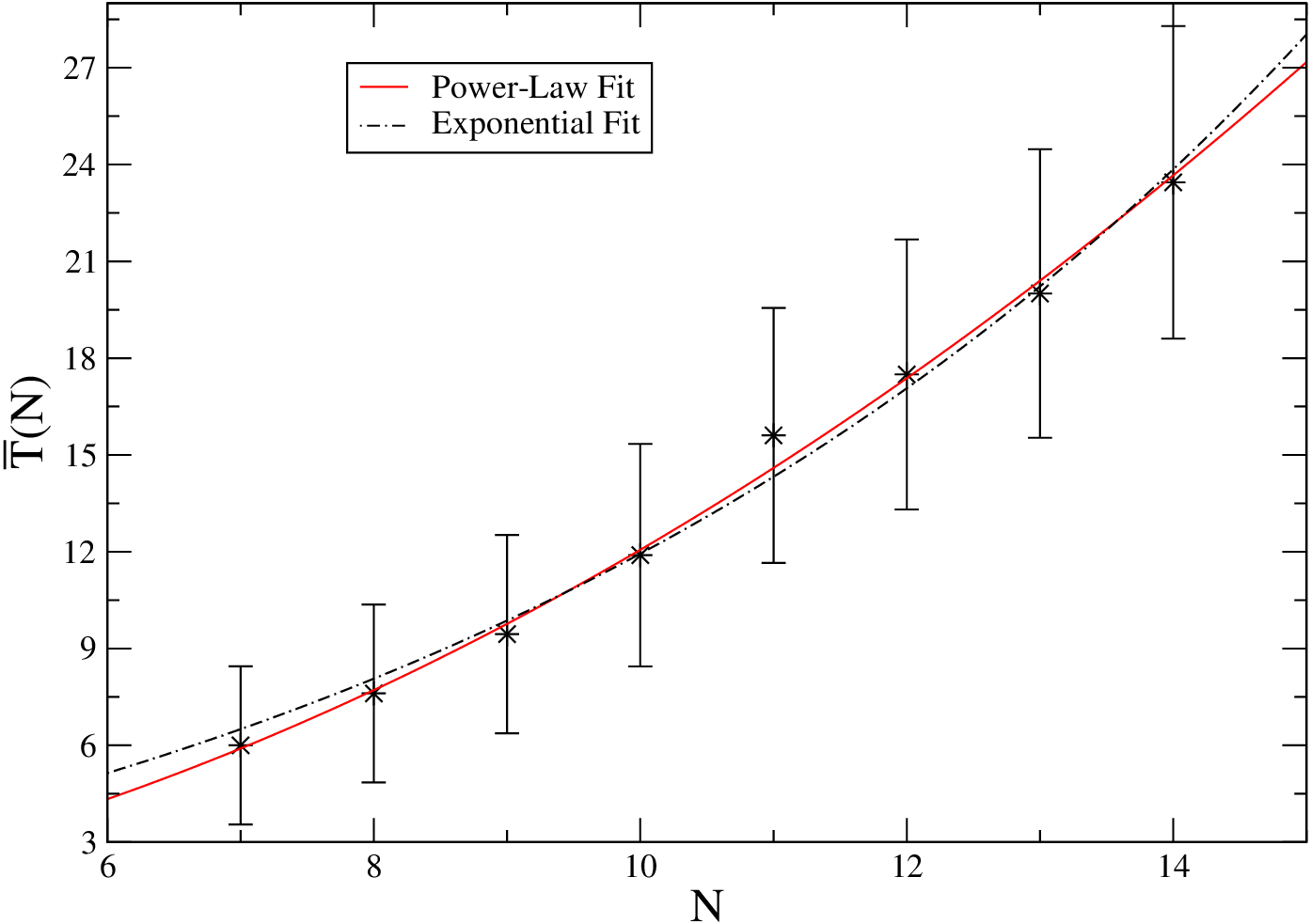}
\vspace{0.1in}
\caption{\label{fig1} Simulation results generated by our QuAdS numerical code 
for the noiseless median runtime $\overline{T}(N)$ (dimensionless units) 
versus the number of bits $N$. The solid line is the best power-law fit 
to the data and the dash-dot line is the best exponential fit. 
Respective values for $a$ and $b$ are given in the text. The error bars give 
95\% confidence limits for each median. Each datapoint is the outcome of 
averaging over 75 USA instances\vspace{0.in}.}
\end{figure}

As a check on the soundness of our own simulation code, we repeated 
the noiseless calculation of Farhi et.~al.\ for $7\leq N\leq 14$. 
Figure~\ref{fig1} shows our results for $\overline{T}(N)$ versus $N$, 
together with a power-law fit to the simulation data: $\overline{T}(N) = 
aN^{b}$. The best-fit parameters are $a=0.11966$ and $b=2.0034$. The 
$\chi^{2}$ value for the fit is $0.092$, and the probability $P(\chi^{2}\geq 
0.092) = 0.9999844$. 
As noted in Section~\ref{sec4}, the closer this probability is to $1$, the
more consistent the data-set is with the fitting function.
We see that a power-law fit provides excellent agreement with our simulation
results, and our exponent $b=2.0034$ is consistent with the quadratic fit of 
Ref.~\cite{fa1}. Figure~\ref{fig1} also includes an exponential fit to the 
noiseless data. The best-fit parameters are: $a=3.89566$; $b=0.140235$; 
$\chi^{2}_{fit} = 0.214$; and $P(\chi^{2}>0.214) = 0.9998121$.
We see that exponential scaling also provides an excellent fit to the
data. We will comment further on this in Sections~\ref{sec4} and \ref{sec5}.
The addition of noise to the simulation of QuAdS causes 
simulation to become impractical when $N \agt 12$. In a couple of cases
we were able to simulate $N= 13\: \mathrm{and}\: 14$ (see Section~\ref{sec4}). 
Our noiseless simulation will be used as a baseline for our simulation of 
QuAdS in the presence of noise. Since the simulations with noise do not go 
beyond $N=14$, we did not extend the noiseless simulations beyond $N=14$. 

\section{\label{sec3} Noisy Quantum Adiabatic Search}

In this section we discuss QuAdS in the presence of noise. Section~\ref{sec3a}
presents our noise model and describes how noise realizations are constructed.
Section~\ref{sec3b} discusses how the noiseless simulation protocol is
modified to allow a study of the noise-averaged performance of QuAdS.

\subsection{\label{sec3a}Noise Model}

We consider noise $\bNi (t)$ that couples to the qubits via the Zeeman
interaction:
\begin{equation}
\label{Hint}
H_{int} = - \sum_{i=1}^{N}\, \gamma_{i}\,\mbox{\boldmath$\sigma_{i}$}
               \cdot\bNi (t) 
                \hspace{0.1in} .
\end{equation}
The qubits are labeled by the index $i$ and the subscript $i$ on the noise
field $\bNi (t)$ and coupling constant $\gamma_{i}$ indicates that they can 
vary from one qubit site to another. We assume the noise is a stationary 
random process. To simplify the presentation of our noise model we initially 
consider a one-component noise field $N(t)$. The presentation is easily 
generalized to three-component spatially varying noise $\bNi (t)$.

The rate at which the noise field $N(t)$ can do work (i.~e.\ noise 
power) is \cite{tho},
\begin{displaymath}
P = N^{2}(t) \hspace{0.1in} ,
\end{displaymath}
and the energy that can be delivered in a time interval $dt$ is,
\begin{displaymath}
dE = N^{2}(t)\, dt \hspace{0.1in} .
\end{displaymath}
We consider power-type noise for which the time-averaged noise power
\begin{equation}
\label{meanpow}
\overline{P} = \lim_{T\rightarrow\infty}\:\frac{1}{T}\,\int_{-T/2}^{T/2}\,
                 N^{2}(t)\, dt 
\end{equation}
is finite. The total noise energy 
\begin{equation}
\label{Etot}
E = \int_{-\infty}^{\infty}\, dt\, N^{2}(t) 
\end{equation}
diverges for this class of noise. The divergence is due to the 
occurrence of an infinite number of noise fluctuations in the time interval 
$-\infty < t < \infty$. The energy of an individual fluctuation is, however, 
finite. 

The mean noise power $\overline{P}$ can be related to the noise correlation
function,
\begin{equation}
\label{corfcn}
\overline{N(t)N(t-s)} \equiv \lim_{T\rightarrow\infty}\,\frac{1}{T}\,
                         \int_{-T/2}^{T/2}\, dy\, N(y)N(y-s) \hspace{0.1in} .
\end{equation}
Comparing eqns.~(\ref{meanpow}) and (\ref{corfcn}) we see that,
\begin{equation}
\label{pow1}
\meanP = \overline{N^{2}(t)} \hspace{0.1in} .
\end{equation}
The Weiner-Khintchine theorem \cite{rie} shows that the noise correlation
function and the power spectral density $S_{N}(f)$ form a Fourier transform
pair: 
\begin{equation}
\label{wkthm}
\overline{N(t)N(t-s)} = \int_{-\infty}^{\infty}\, df\, \specdenN\,
                           e^{-2\pi ifs} \hspace{0.1in} .
\end{equation}
Thus, it follows from eqns.~(\ref{pow1}) and (\ref{wkthm}) that
\begin{equation}
\label{powspec}
\meanP = \int_{-\infty}^{\infty}\, df\, \specdenN \hspace{0.1in} ,
\end{equation}
which identifies $\specdenN$ as the mean noise power available in the
frequency interval ($f$, $f+df$). 

Having reviewed these basic facts about stationary random processes, we are
now in a position to describe our noise model and to write out the basic
relations that are necessary to construct an individual realization of the 
noise.

The noise $N(t)$ is produced by a sequence of randomly occurring noise 
fluctuations $F(t)$. The fluctuations: (1) occur independently of each other 
at average rate $\meann\,$ per unit time; and (2) have a peak value 
$x$ which is Gaussian distributed with mean $\overline{x} = 0$, variance 
$\overline{x^{2}}= \sigma^{2}$, and temporal width $2\tau$, where $\tau$ is 
the thermal relaxation time. The bandwidth of $F(t)$ is $\Delta\omega \sim 1/
2\tau$. Thus,
\begin{displaymath}
N(t) = \sum_{i}\, F(t-t_{i}) \hspace{0.1in} ,
\end{displaymath}
where $i$ labels the noise fluctuations, and $t_{i}$ specifies the center of
the $i$th fluctuation. The mean number of fluctuations 
$\overline{\mathcal{N}}_{\hspace{-0.25em}f}$ occurring in the 
time interval $[0,\: T]$ is $\overline{\mathcal{N}}_{\hspace{-0.25em}f} = 
\meann \: T$. 
It is well-known that for noise with these properties, the actual number of 
fluctuations $n$ that occur in a time $T$ is governed by the Poisson 
distribution \cite{drk}:
\begin{displaymath}
p(n) = \frac{(\overline{\mathcal{N}}_{\hspace{-0.25em}f})^{n}}{n!}\, 
         e^{-\overline{\mathcal{N}}_{\hspace{-0.25em}f}}
          \hspace{0.1in} .
\end{displaymath}
The energy present in a single fluctuation is:
\begin{equation}
\label{flucE}
\varepsilon = \int_{-\infty}^{\infty}\, F^{2}(t) \, dt \hspace{0.1in} .
\end{equation}
Let $F(t) = x\, h(t)$, where $h(t)$ is any convenient function of finite
support with normalization
\begin{displaymath}
\int_{-\infty}^{\infty}\, dt\, h^{2}(t) = 2\tau \hspace{0.1in} .
\end{displaymath}
As mentioned above, $x$ is Gaussian distributed with mean 
$\overline{x} = 0$ and $\overline{x^{2}}=\sigma^{2}$. From 
eqn.~(\ref{flucE}), $\varepsilon = 2x^{2}\, \tau$, and the mean energy per 
fluctuation $\overline{\varepsilon}$ is,
\begin{equation}
\label{meanfE}
\overline{\varepsilon} = 2\,\overline{x^{2}}\,\tau = 2\sigma^{2}\,\tau
  \hspace{0.1in} .
\end{equation}
From Campbell's theorem \cite{ric}, the power spectral density for $N(t)$ is
\begin{equation}
\label{campthm}
\specdenN = \meann\, |g(f)|^{2} \hspace{0.1in} ,
\end{equation}
where $g(f)$ is the Fourier transform of the fluctuation profile
$F(t)$. Thus, using eqns.~(\ref{powspec}), (\ref{campthm}), and Paresval's
theorem gives,
\begin{equation}
\meanP = \meann\,\int_{-\infty}^{\infty}\, dt\, F^{2}(t) \hspace{0.1in} .
\end{equation}
Finally, using eqns.~(\ref{flucE}) and (\ref{meanfE}) gives,
\begin{equation}
\label{finlmeanP}
\meanP = 2\,\meann\,\sigma^{2}\,\tau \hspace{0.1in} .
\end{equation}
Thus we see that our noise model is characterized by any 3 of the parameters 
$\meanP$, $\meann$, $\sigma^{2}$, and $\tau$.

The numerical simulation constructs a realization of the noise as follows.
We sample a positive integer $N_{f}$ according to the Poisson distribution 
with mean $\overline{\mathcal{N}}_{\hspace{-0.25em}f} = \meann\: T$, where $T$ 
is the duration
of the search. $N_{f}$ is the number of fluctuations present in the noise 
realization. The noise model assumes these fluctuations occur independently 
with probability $dp_{f} = (1/T)dt$. We sample $N_{f}$ numbers $t_{i}$
($i= 1, \cdots , N_{f}$) from the interval $(0,\: T)$. The $t_{i}$ correspond 
to the temporal centers of the $N_{f}$ fluctuations. For simplicity, we've 
assumed that the fluctuation profile $h(t)$ is a square pulse. We next carry 
out $N_{f}$ samples $x_{i}$ ($i = 1, \cdots , N_{f}$) of a Gaussian 
distribution with mean $\overline{x}_{i}=0$ and variance $\overline{x^{2}_{i}} 
= \sigma^{2}$. Here $x_{i}$ is the peak value of the $i$-th fluctuation. 
These sample results produce the noise realization $N(t)$:
\begin{equation}
\label{noyzrel}
N(t) = \sum_{i=1}^{N_{f}}\, x_{i}\,\left[\,
             \frac{\sgn (t - t_{il}) - \sgn (t-t_{ir})}{2}
           \,\right] \hspace{0.1in} , 
\end{equation}
where $t_{il} = t_{i} -\tau$, and $t_{ir} = t_{i}+\tau$. We shall need to 
produce noise realizations with arbitrary mean noise power $\meanP$. We do 
this by the following normalization procedure. First we calculate the mean
noise power $\overline{\mathcal{P}}$ of the noise realization $N(t)$ just
produced:
\begin{equation}
\label{rawpow}
\overline{\mathcal{P}} = \frac{1}{T}\,\int_{0}^{T}\, dt \, N^{2}(t) 
                           \hspace{0.1in} .
\end{equation}
Then, if the desired value for the noise power is $\meanP$, we rescale
$N(t)$ in eqn.~(\ref{noyzrel}) so that $N(t)\rightarrow n(t) \equiv 
\sqrt{\meanP / \overline{\mathcal{P}}}\, N(t)$. The result is a noise 
realization $n(t)$ with mean noise power $\meanP$. The simulation takes as 
inputs the mean noise power $\meanP$, the variance 
$\overline{x_{i}^{2}} = \sigma^{2}$, and the thermal relaxation time $\tau$.
The fluctuation rate $\meann$ then follows from eqn.~(\ref{finlmeanP}):
$\meann = \meanP /(2\sigma^{2}\tau )$. This procedure is used to produce 4
types of noise that are characterized by the polarization, or direction along 
which the noise $\bNi (t)$ fluctuates: (i) $\hat{\mathbf{x}}$;
(ii) $\hat{\mathbf{y}}$; (iii) $\hat{\mathbf{z}}$; and (iv) all 3 directions 
simultaneously. We shall refer to these noise polarization types as x-type;
y-type; z-type; and 3-type noise, respectively. The noise $\bNi (t)$ is then 
introduced into eqn.~(\ref{Hint}) and the full Hamiltonian $\mathcal{H}(t)$ is 
the sum of $H(t)$ (eqn.~(\ref{timeHam})) and $H_{int}(t)$ (eqn.~(\ref{Hint})):  
\begin{equation}
\label{totHam}
\mathcal{H}(t) = H(t) + H_{int}(t) \hspace{0.1in} .
\end{equation}
$\mathcal{H}(t)$ drives the Schrodinger dynamics of QuAdS in the presence
of the noise realization $\bNi (t)$. We now go on to explain how the
noiseless simulation protocol of Ref.~\cite{fa1} is extended to include the 
effects of noise.

\subsection{\label{sec3b}Noisy Simulation Protocol}

We would like to compare how the 4 different noise types introduced above 
impact the performance of QuAdS. To that end, we determine the scaling 
relation for the noise-averaged median runtime $\langle T(N)\rangle$ versus 
$N$ for each type of noise. Just as in the noiseless protocol, for each 
value of $N$, we generate 75 USA instances of EC3. QuAdS, in the presence of
each of the 4 noise types, is then applied to the \textit{same\/} 75 USA 
instances. This allows an apples-to-apples comparison of how the different 
noise types affect QuAdS performance. For each noise type, and each USA 
instance, we generate 10 noise realizations and use the Hamiltonian 
$\mathcal{H}(t)$ to find 
the 10 corresponding QuAdS runtimes $T_{i}^{j}(N)$. Here $j$ labels the 
noise realizations ($j = 1, \cdots , 10$), and $i$ labels the USA instances 
($i = 1, \cdots , 75$). Thus, for each value of $N$, and each noise type, 750 
runtimes are found. The ensemble $\mathcal{E}_{N}$ of 750 runtimes 
$T_{i}^{j}(N)$ allows us to estimate the statistical effects of each noise 
type on QuAdS performance over the full set of 75 USA instances. The 
noise-averaged median runtime $\langle T(N)\rangle$ is then identified with 
the median runtime calculated from the noise ensemble $\mathcal{E}_{N}$. 
We then fit the simulation results using both power-law scaling $\langle T(N)
\rangle = aN^{b}$ and exponential scaling $\langle T(N) \rangle = a\left[
\exp (bN) -1\right]$, and calculated the $\chi^{2}$ for each type of fit. We 
determine 2 such fits for each of the 4 noise types. For each fit-type, we 
compared the respective scaling exponents to get a quantitative measure of 
which of the 4 noise types most adversely affects QuAdS. 
As will be discussed in Section~\ref{sec4}, scaling
curves will be determined for each of the 4 noise types for 3 different
noise power levels. As pointed out at the end of Section~\ref{sec2c}, most
simulations will be restricted to $7 \leq N\leq 12$, though for x-type noise,
the upper limit could be extended to $N=13$ and $14$. It is important to 
recognize that each point on a given scaling curve is based on 750 runtimes, 
and so each scaling curve with $7\leq N\leq 12$ ($7\leq N\leq 14$)
is distilled from 4500 (6000) runtimes. Each runtime $T_{i}^{j}(N)$ is 
itself the result of a hunting procedure that requires, on average, 
5 integrations of the Schrodinger equation. Thus a single scaling curve 
corresponds to approximately 22,500 (30,000) integrations of the Schrodinger 
equation. As mentioned above, we generate 4 such curves for each noise power 
level, and we present data for 3 power levels. Putting all this together,
we see that the simulation results we present in this paper are the outcome of 
approximately 300,000 integrations of the Schrodinger equation. This work was 
done, initially on our own 16-node Beowulf cluster, and later on the TeraGrid 
cluster which was accessed through the National Center for Supercomputing 
Applications (NCSA) in Urbana, Illinois.

\section{\label{sec4} Simulation Results}

In this Section we present our simulation results for noisy QuAdS. The noise 
model parameters (see Section~\ref{sec3}) were chosen to be 
$\sigma = 0.2$; $\tau = 1$; and $\meanP = 0.001,\: 0.003,\: 0.005$. This 
produced noise that was strong enough to affect QuAdS performance, though not 
so strong as to make large-scale simulation impractical. Both $H_{i}$ and 
$H_{p}$ have energy--level spacing $\Delta E\sim 1$. For our choice of $\tau$, 
$\Delta E \sim 2/\tau$, so that our noise bandwidth matches the natural
resonance energies of the quantum computer. We are thus probing QuAdS in a
noise regime that complements that studied in Ref.~\cite{rc2}. Further
discussion of our choice of noise parameters will be given in 
Section~\ref{sec5}. As this was our first large-scale simulation of noisy 
QuAdS, we decided to simplify the noise model by restricting the noise field 
and coupling constants to be the same for each qubit: $\bNi (t) = \bfN (t)$
and $\gamma_{i}=1$. We plan to lift this restriction in our next set of 
simulations. We also plan to parallelize the hunting procedure which should 
substantially speed up the code, allowing us to explore larger values of mean 
noise power $\meanP$. To get a sense of how long the following 
simulation took in real time, note that producing one scaling curve at 
$\meanP = 0.005$ took approximately 3 weeks on our 16-node Beowulf 
cluster. The same simulation took approximately $72$--$96$ hours on the 
TeraGrid cluster at NCSA. 

\begin{figure}[!]
\includegraphics[scale=0.325]{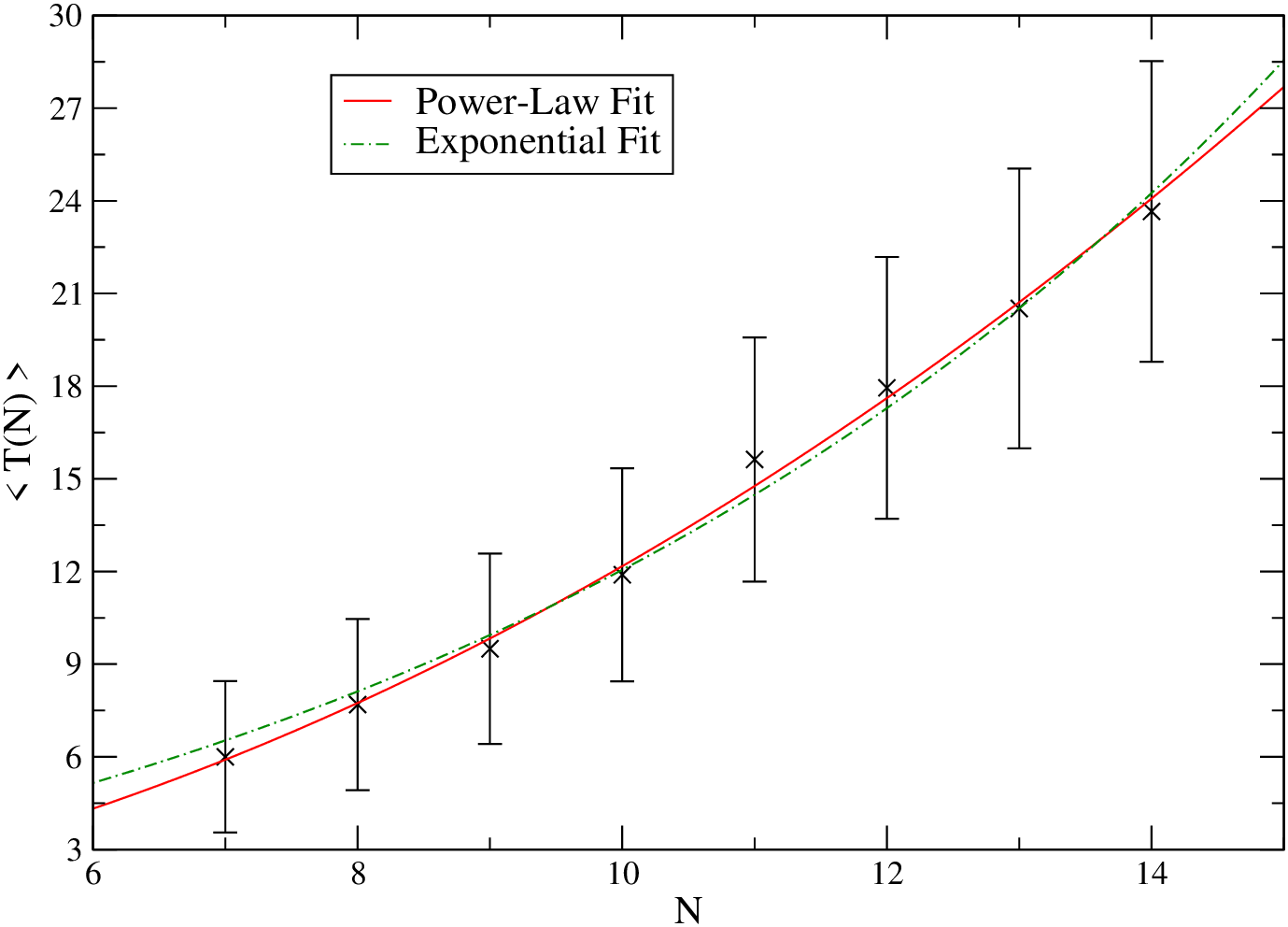}
\vspace{0.1in}
\caption{\label{fig2}Simulation results for QuAdS with noise polarized 
along $\hat{\mathbf{x}}$ and mean noise power $\overline{P}=0.001$. Plotted 
are the noise-averaged median runtime $\langle T(N)\rangle$ (dimensionless 
units) versus the number of bits $N$. The solid line is the best power-law 
fit to the data and the dash-dot line is the best exponential fit to the
data. The error bars give 95\% confidence limits for each median. Each 
datapoint is the outcome of averaging over 75 USA instances and 10 noise 
realizations per USA instance\vspace{0.in}.}
\end{figure}

\begin{figure}[!]
\vspace{0.20in}
\includegraphics[scale=0.325]{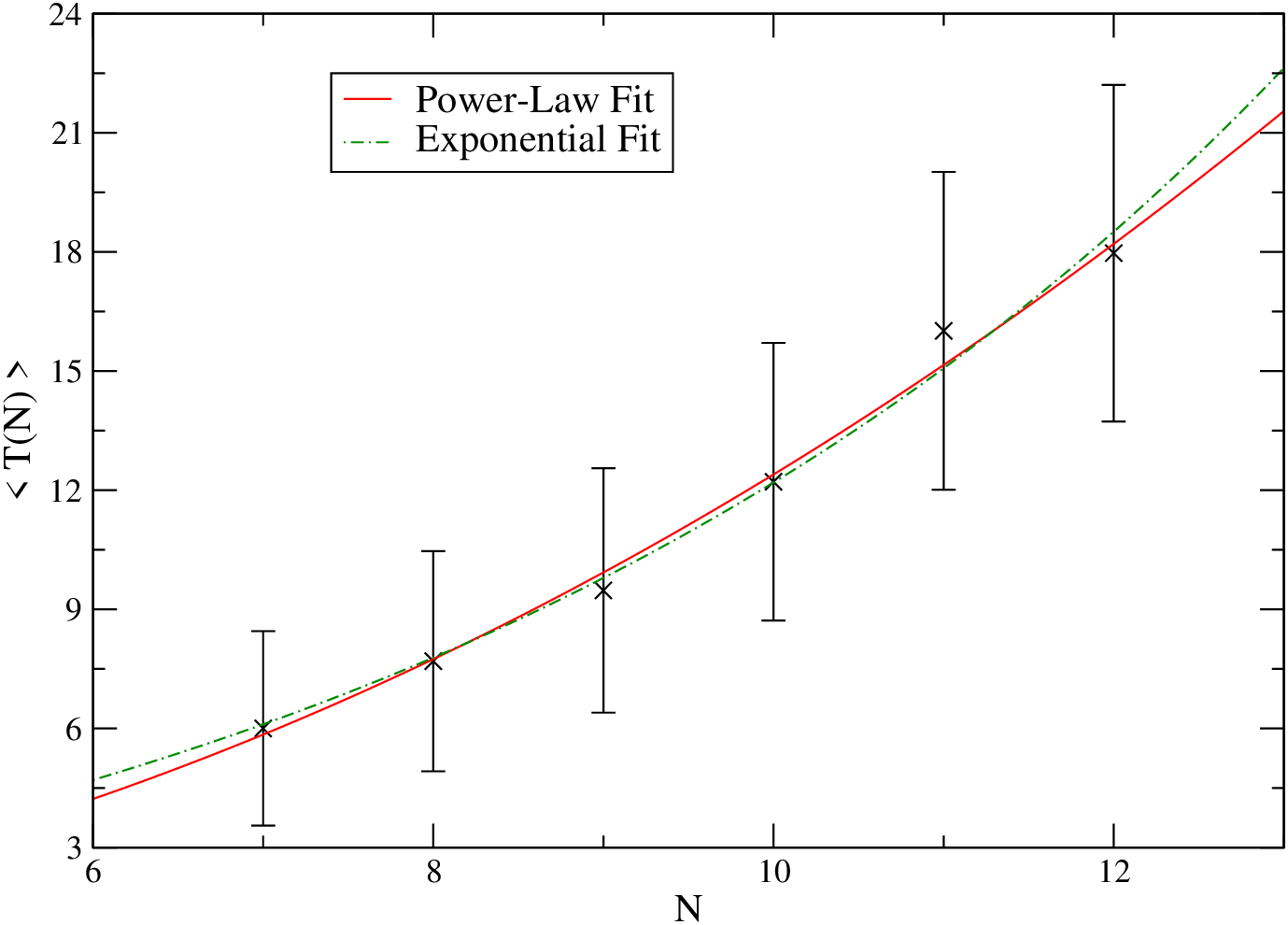}
\vspace{0.1in}
\caption{\label{fig3}Simulation results for QuAdS with noise polarized 
along $\hat{\mathbf{y}}$ and mean noise power $\overline{P}=0.001$. Plotted 
are the noise-averaged median runtime $\langle T(N)\rangle$ (dimensionless 
units) versus the number of bits $N$. The solid line is the best power-law 
fit to the data and the dash-dot line is the best exponential fit to the
data. The error bars give 95\% confidence limits for each median. Each 
datapoint is the outcome of averaging over 75 USA instances and 10 noise 
realizations per USA instance\vspace{0.in}.}
\end{figure}

\begin{figure}[!]
\vspace{0.20in}
\includegraphics[scale=0.325]{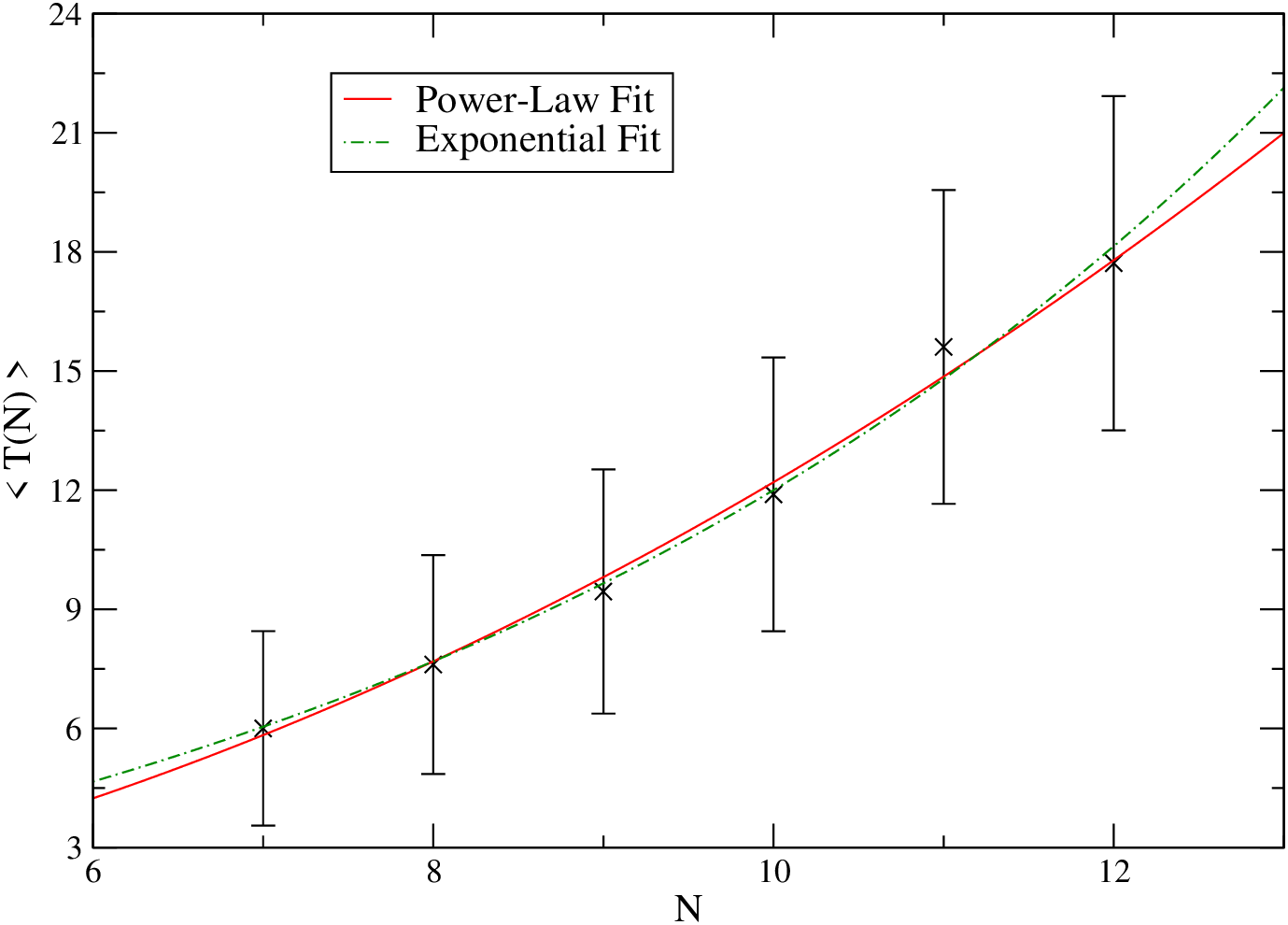}
\vspace{0.1in}
\caption{\label{fig4}Simulation results for QuAdS with noise polarized 
along $\hat{\mathbf{z}}$ and mean noise power $\overline{P}=0.001$. Plotted 
are the noise-averaged median runtime $\langle T(N)\rangle$ (dimensionless 
units) versus the number of bits $N$. The solid line is the best power-law 
fit to the data and the dash-dot line is the best exponential fit to the
data. The error bars give 95\% confidence limits for each median. Each 
datapoint is the outcome of averaging over 75 USA instances and 10 noise 
realizations per USA instance\vspace{0.in}.}
\end{figure}

\begin{figure}[!]
\includegraphics[scale=0.325]{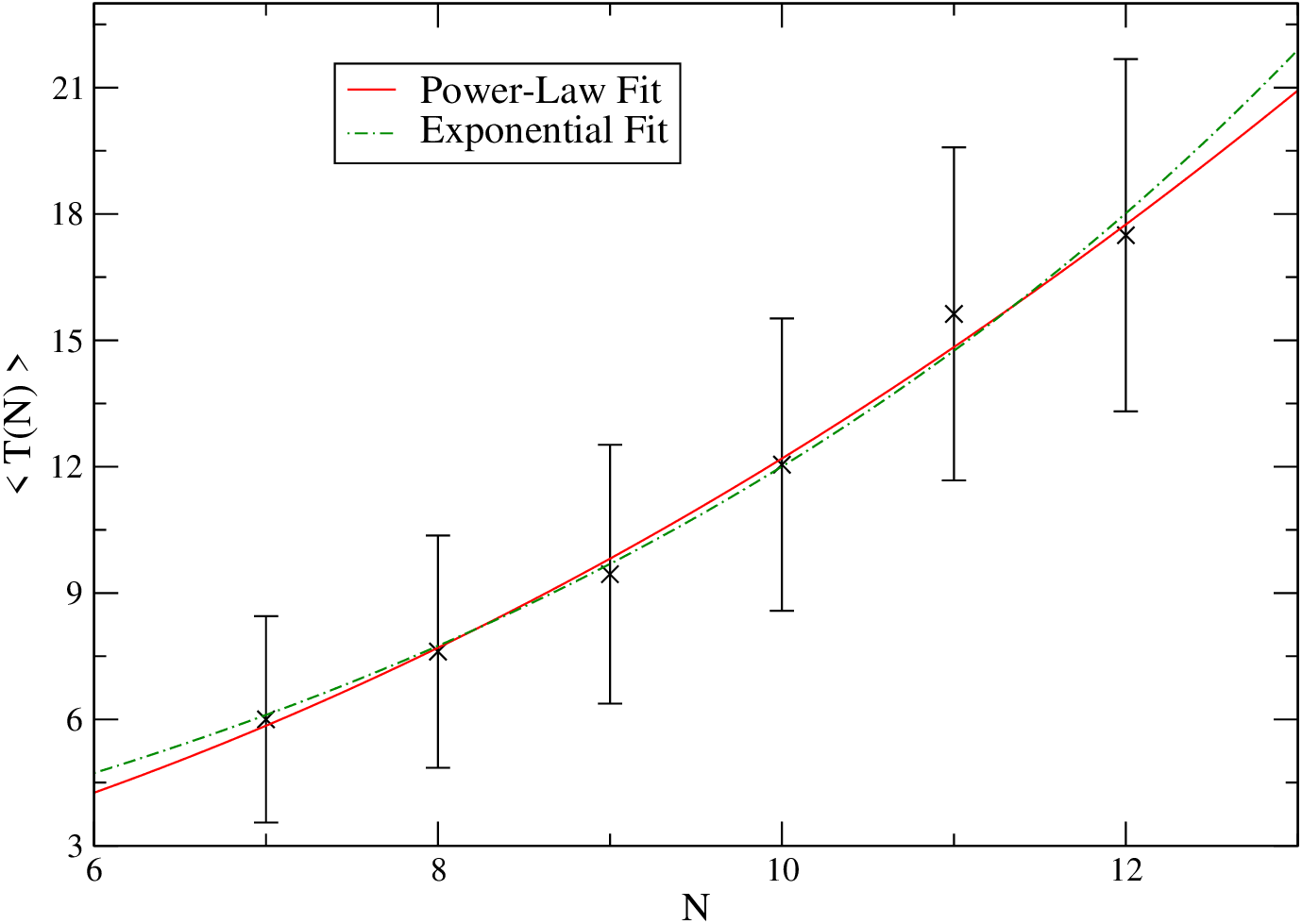}
\vspace{0.1in}
\caption{\label{fig5}Simulation results for QuAdS with noise polarization
along all 3 directions and mean noise power $\overline{P}=0.001$. Plotted 
are the noise-averaged median runtime $\langle T(N)\rangle$ (dimensionless 
units) versus the number of bits $N$. The solid line is the best power-law 
fit to the data and the dash-dot line is the best exponential fit to the
data. The error bars give 95\% confidence limits for each median. Each 
datapoint is the outcome of averaging over 75 USA instances and 10 noise 
realizations per USA instance\vspace{0.in}.}
\end{figure}

\begin{figure}[!]
\vspace{0.20in}
\includegraphics[scale=0.325]{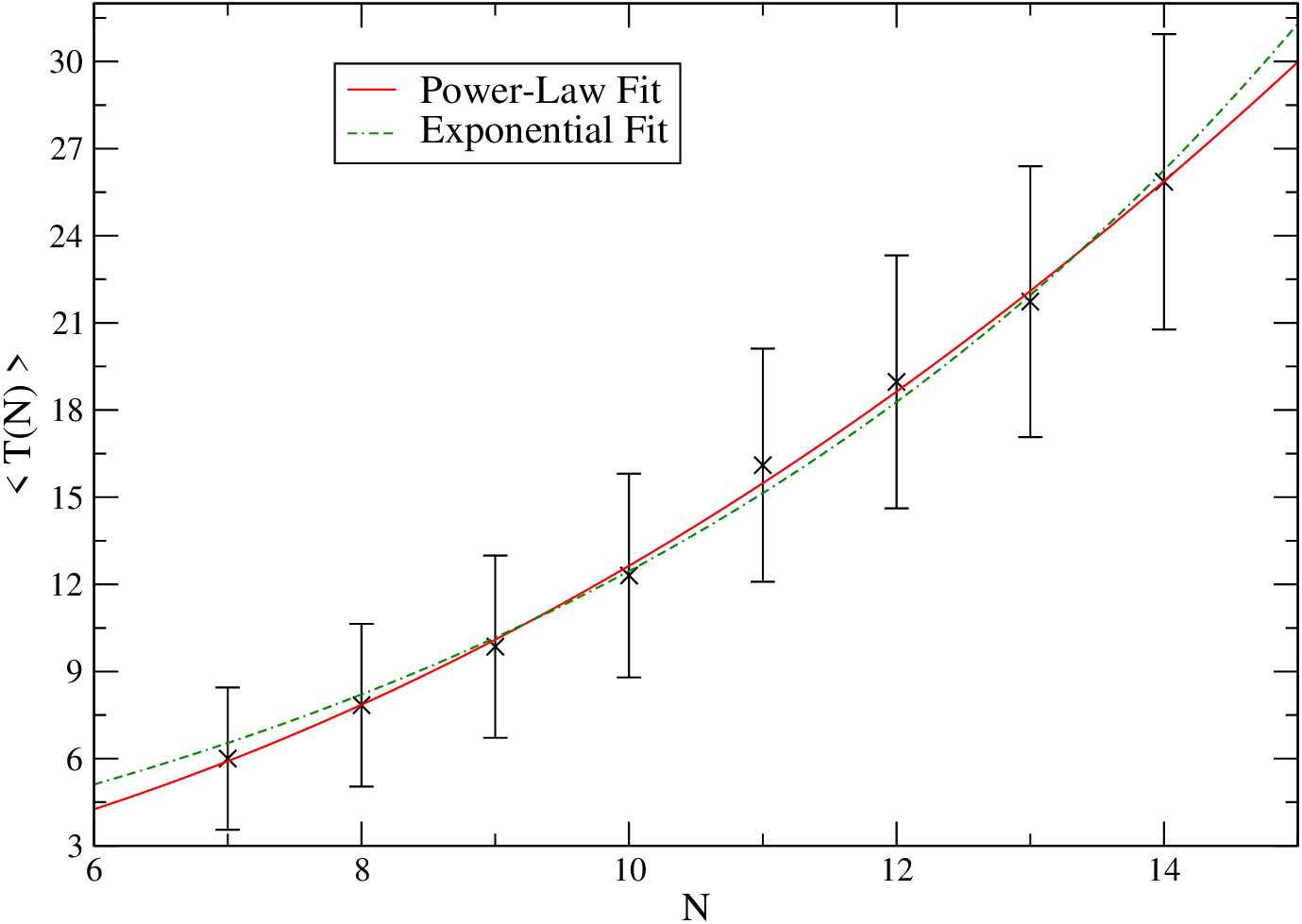}
\vspace{0.1in}
\caption{\label{fig6}Simulation results for QuAdS with noise polarized 
along $\hat{\mathbf{x}}$ and mean noise power $\overline{P}=0.003$. Plotted 
are the noise-averaged median runtime $\langle T(N)\rangle$ (dimensionless 
units) versus the number of bits $N$. The solid line is the best power-law 
fit to the data and the dash-dot line is the best exponential fit to the
data. The error bars give 95\% confidence limits for each median. Each 
datapoint is the outcome of averaging over 75 USA instances and 10 noise 
realizations per USA instance\vspace{0.in}.}
\end{figure}

\begin{figure}[!]
\vspace{0.20in}
\includegraphics[scale=0.325]{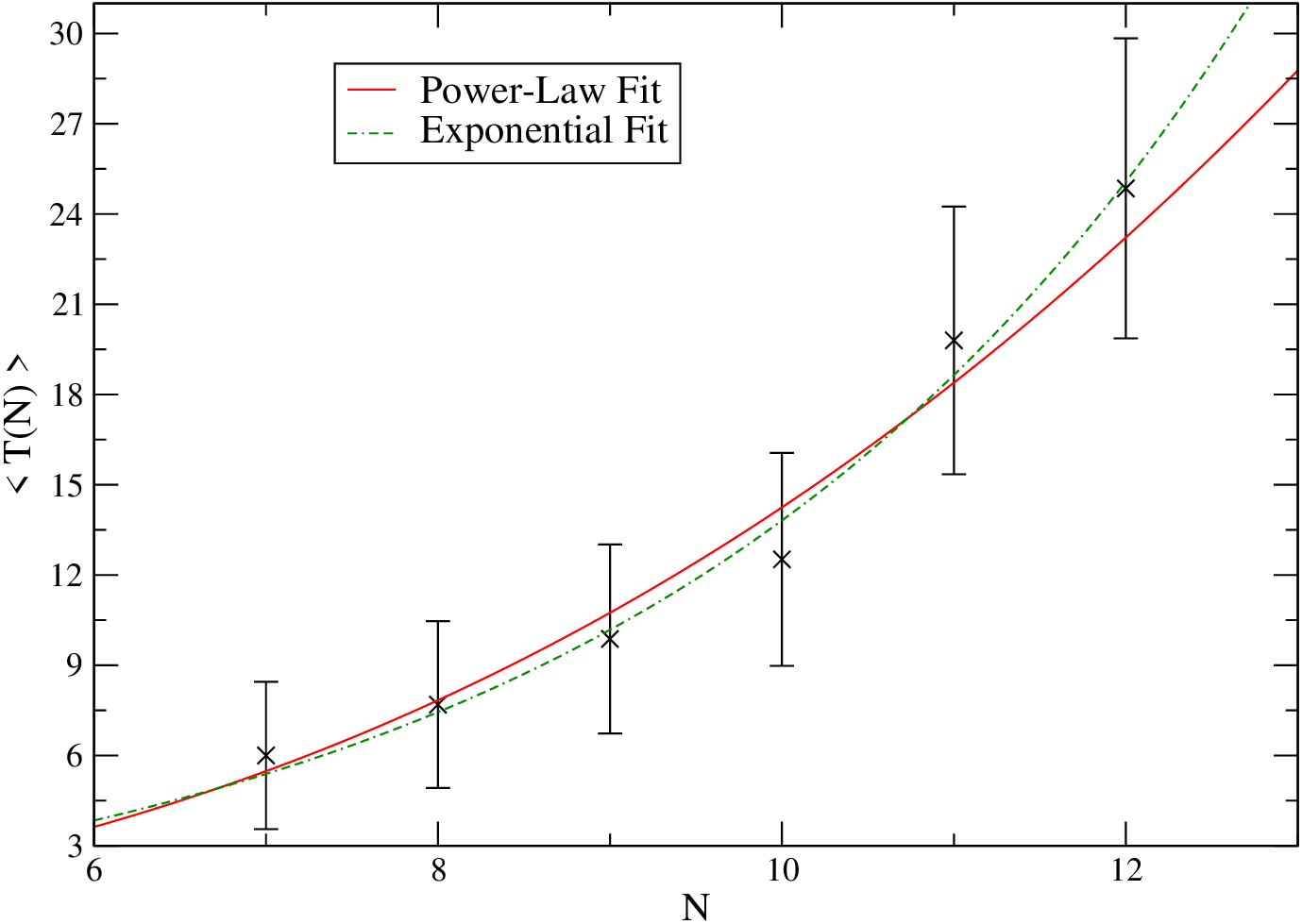}
\vspace{0.1in}
\caption{\label{fig7}Simulation results for QuAdS with noise polarized 
along $\hat{\mathbf{y}}$ and mean noise power $\overline{P}=0.003$. Plotted 
are the noise-averaged median runtime $\langle T(N)\rangle$ (dimensionless 
units) versus the number of bits $N$. The solid line is the best power-law 
fit to the data and the dash-dot line is the best exponential fit to the
data. The error bars give 95\% confidence limits for each median. Each 
datapoint is the outcome of averaging over 75 USA instances and 10 noise 
realizations per USA instance\vspace{0.in}.}
\end{figure}

\begin{figure}[!]
\vspace{0.25in}
\includegraphics[scale=0.325]{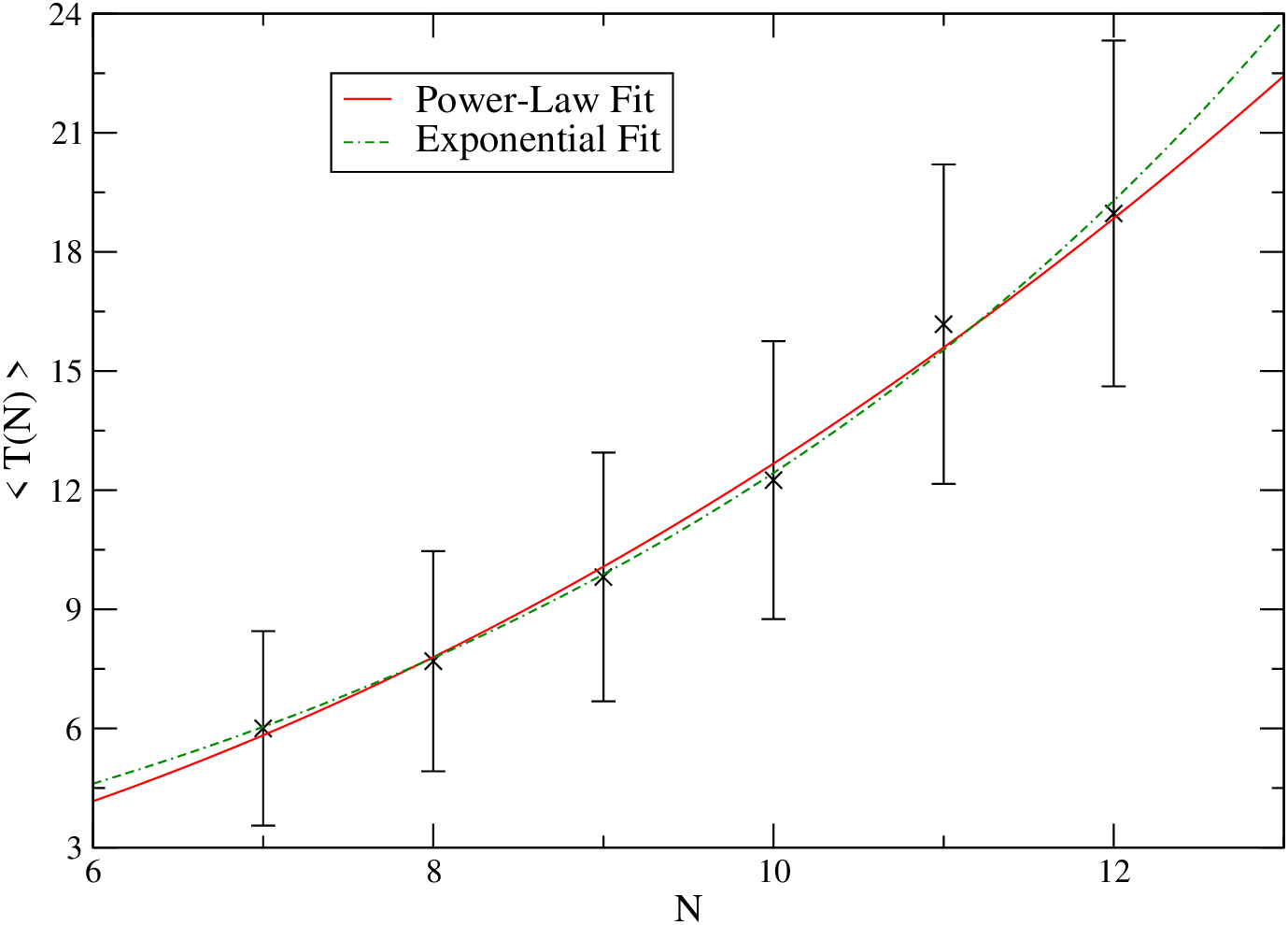}
\vspace{0.1in}
\caption{\label{fig8}Simulation results for QuAdS with noise polarized 
along $\hat{\mathbf{z}}$ and mean noise power $\overline{P}=0.003$. Plotted 
are the noise-averaged median runtime $\langle T(N)\rangle$ (dimensionless 
units) versus the number of bits $N$. The solid line is the best power-law 
fit to the data and the dash-dot line is the best exponential fit to the
data. The error bars give 95\% confidence limits for each median. Each 
datapoint is the outcome of averaging over 75 USA instances and 10 noise 
realizations per USA instance\vspace{0.in}.}
\end{figure}

\begin{figure}[!]
\vspace{0.20in}
\includegraphics[scale=0.325]{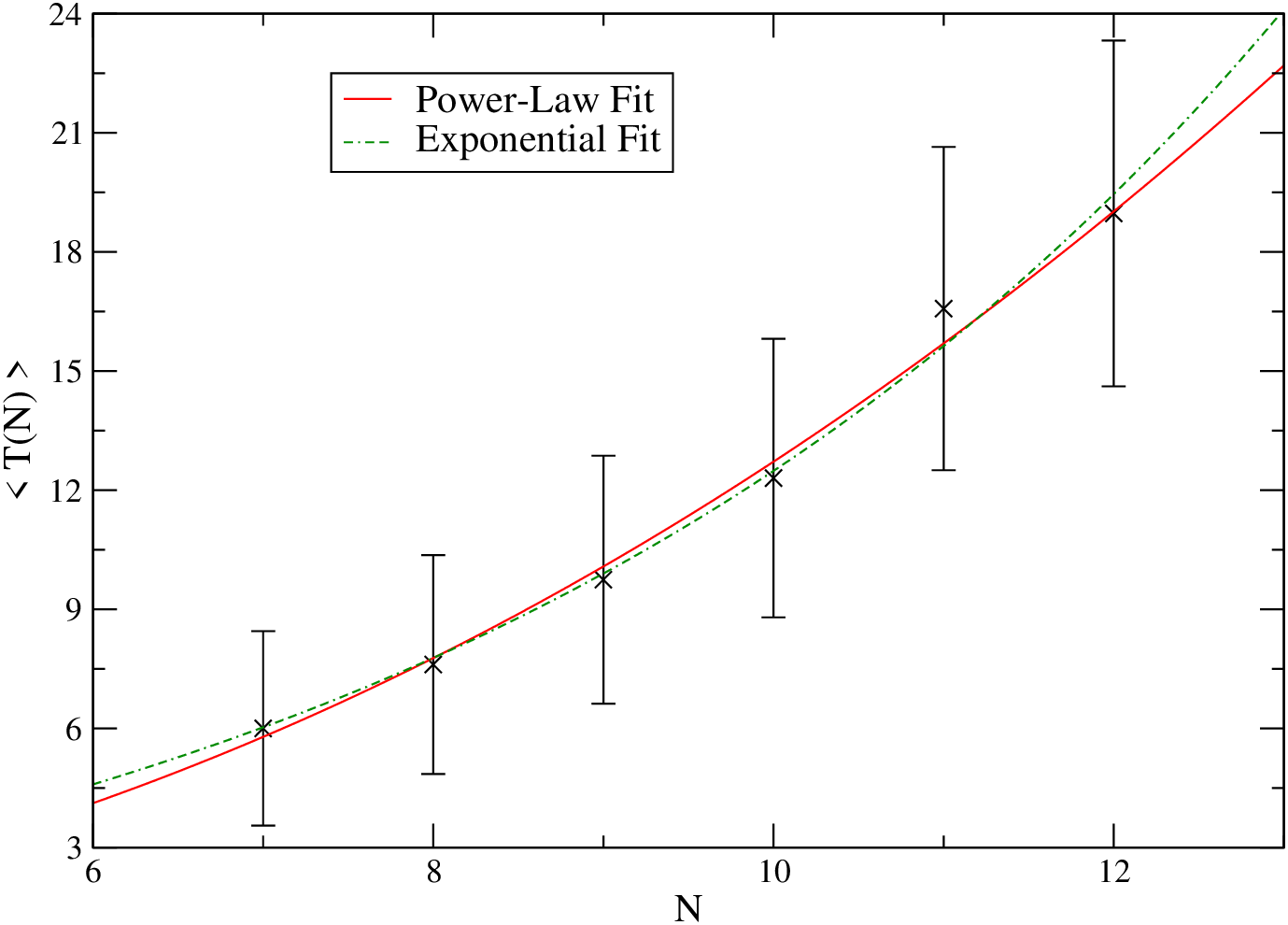}
\vspace{0.1in}
\caption{\label{fig9}Simulation results for QuAdS with noise polarization 
along all 3 directions and mean noise power $\overline{P}=0.003$. Plotted 
are the noise-averaged median runtime $\langle T(N)\rangle$ (dimensionless 
units) versus the number of bits $N$. The solid line is the best power-law 
fit to the data and the dash-dot line is the best exponential fit to the
data. The error bars give 95\% confidence limits for each median. Each 
datapoint is the outcome of averaging over 75 USA instances and 10 noise 
realizations per USA instance\vspace{0.in}.}
\end{figure}

\begin{figure}[!]
\vspace{0.25in}
\includegraphics[scale=0.325]{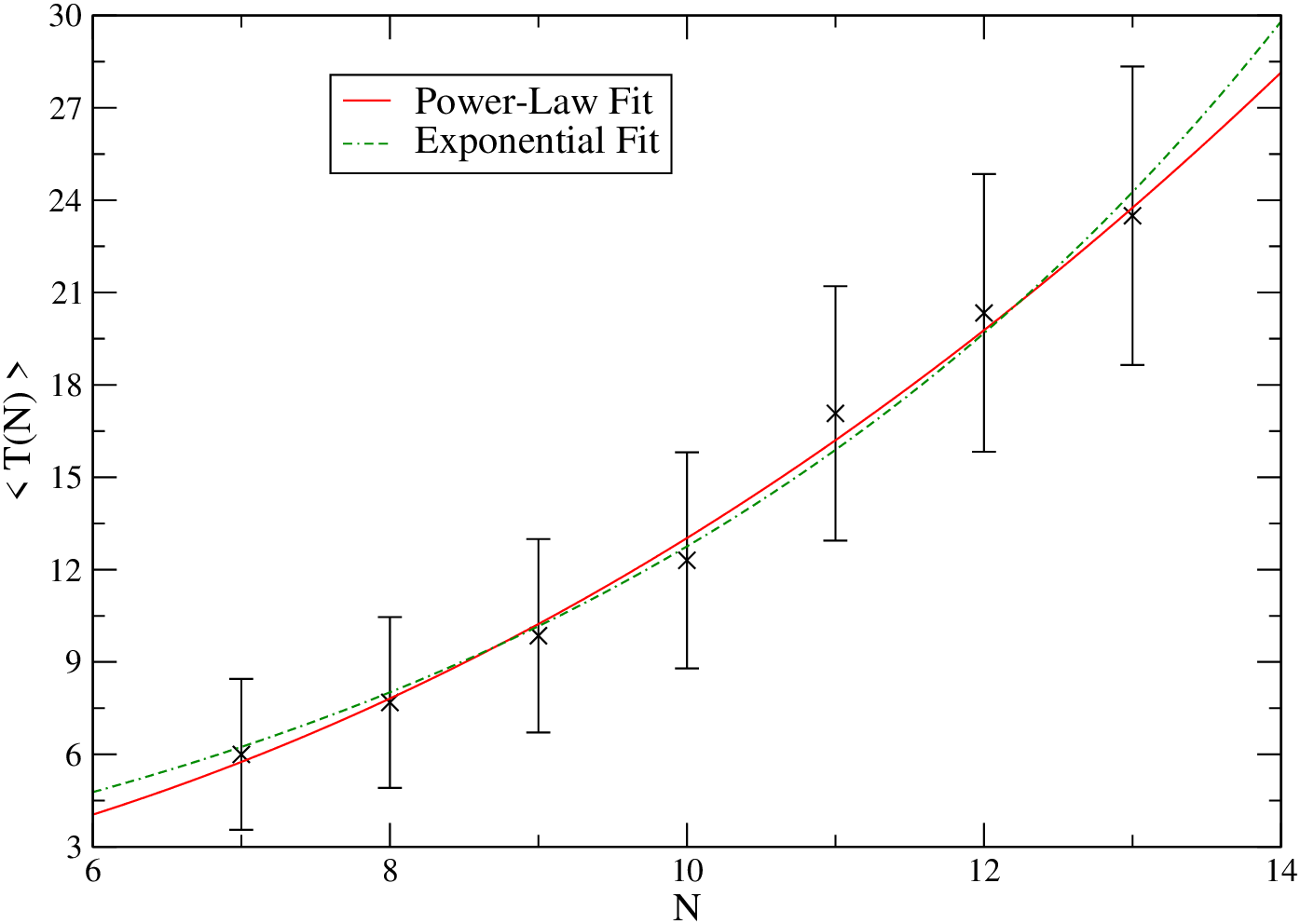}
\vspace{0.1in}
\caption{\label{fig10}Simulation results for QuAdS with noise polarized 
along $\hat{\mathbf{x}}$ and mean noise power $\overline{P}=0.005$. Plotted 
are the noise-averaged median runtime $\langle T(N)\rangle$ (dimensionless 
units) versus the number of bits $N$. The solid line is the best power-law 
fit to the data and the dash-dot line is the best exponential fit to the
data. The error bars give 95\% confidence limits for each median. Each 
datapoint is the outcome of averaging over 75 USA instances and 10 noise 
realizations per USA instance\vspace{0.in}.}
\end{figure}

\begin{figure}[!]
\vspace{0.2in}
\includegraphics[scale=0.325]{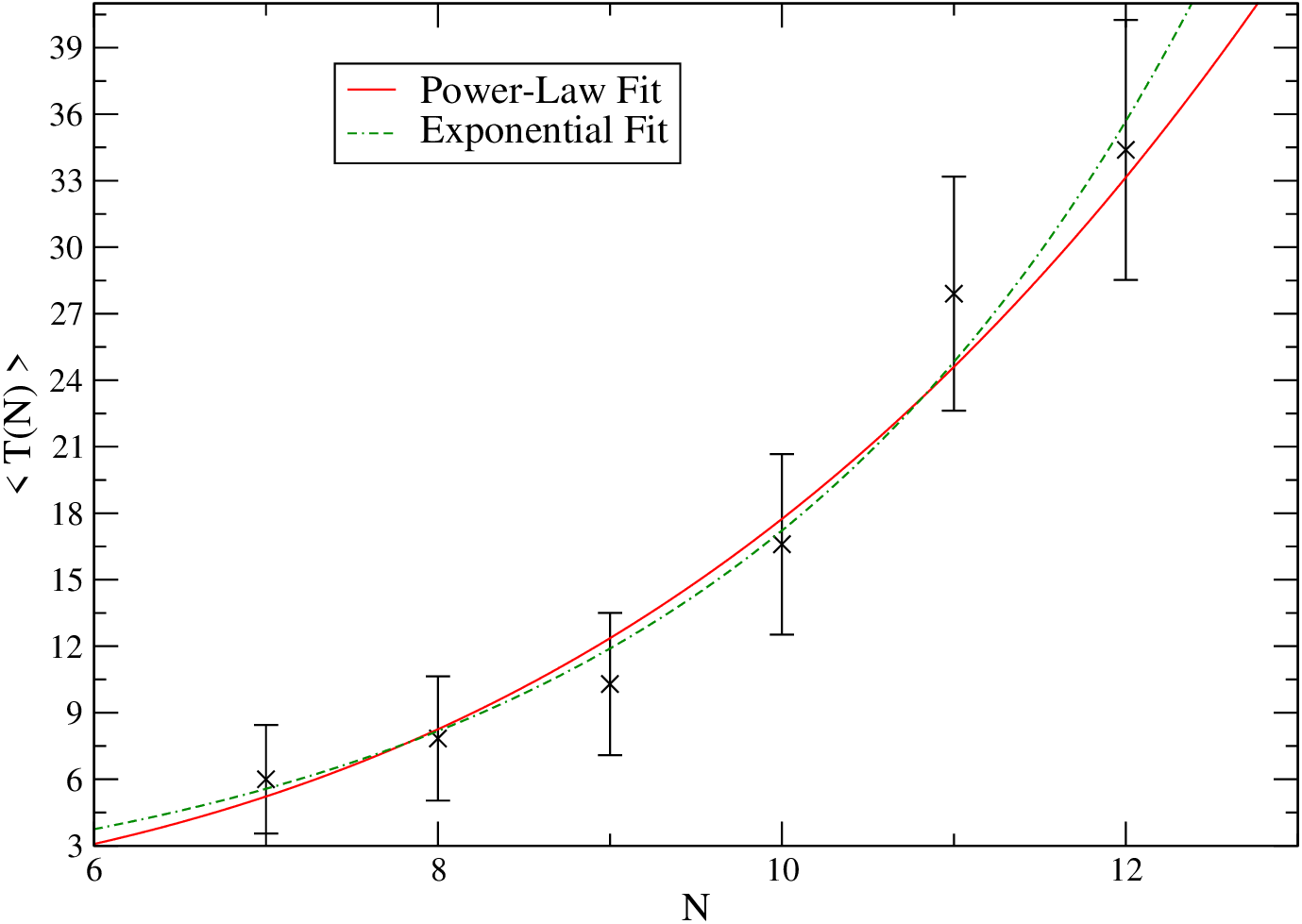}
\vspace{0.1in}
\caption{\label{fig11}Simulation results for QuAdS with noise polarized 
along $\hat{\mathbf{y}}$ and mean noise power $\overline{P}=0.005$. Plotted 
are the noise-averaged median runtime $\langle T(N)\rangle$ (dimensionless 
units) versus the number of bits $N$. The solid line is the best power-law 
fit to the data and the dash-dot line is the best exponential fit to the
data. The error bars give 95\% confidence limits for each median. Each 
datapoint is the outcome of averaging over 75 USA instances and 10 noise 
realizations per USA instance\vspace{0.in}.}
\end{figure}

\begin{figure}[!]
\vspace{0.20in}
\includegraphics[scale=0.325]{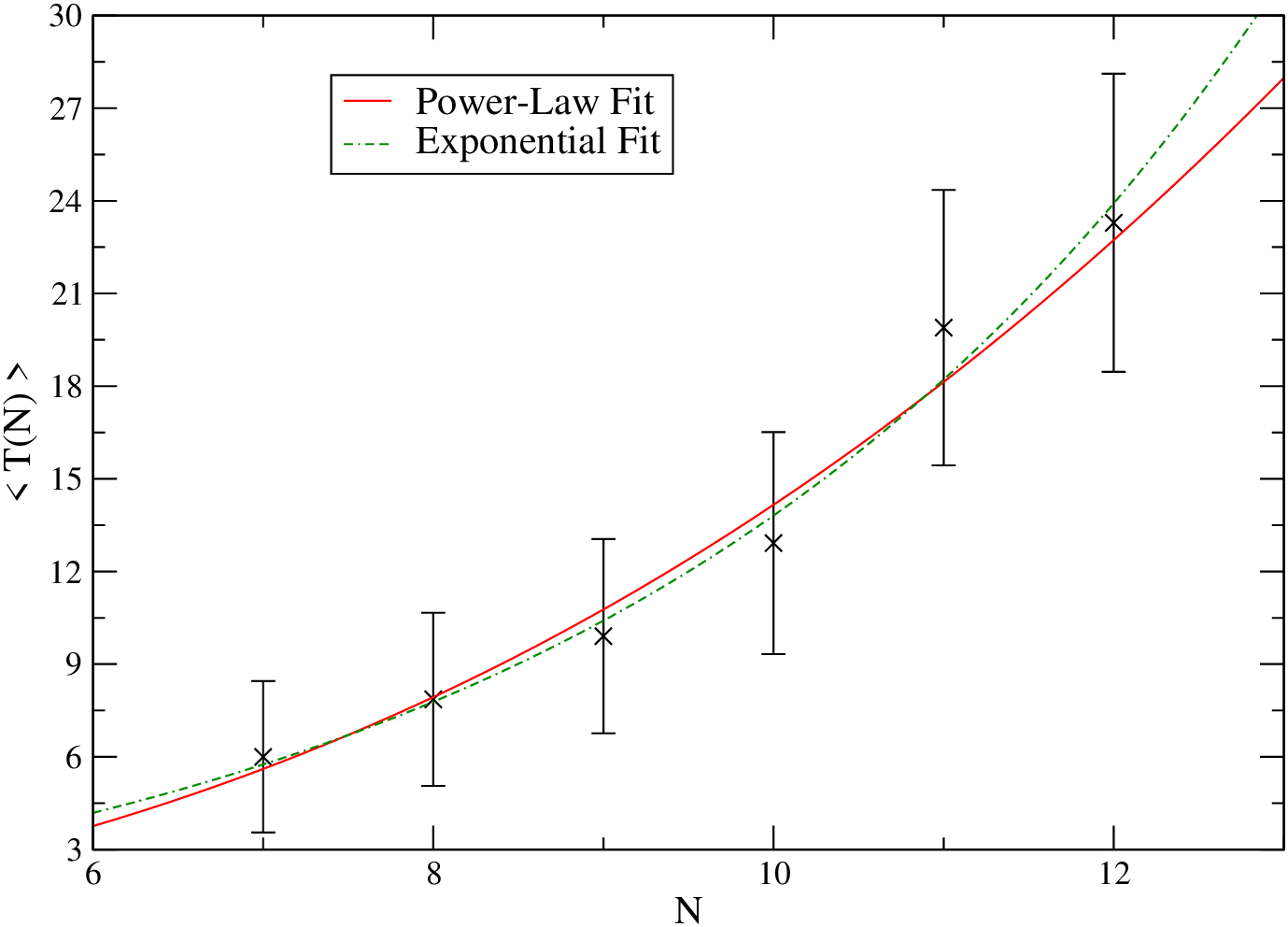}
\vspace{0.1in}
\caption{\label{fig12}Simulation results for QuAdS with noise polarized 
along $\hat{\mathbf{z}}$ and mean noise power $\overline{P}=0.005$. Plotted 
are the noise-averaged median runtime $\langle T(N)\rangle$ (dimensionless 
units) versus the number of bits $N$. The solid line is the best power-law 
fit to the data and the dash-dot line is the best exponential fit to the
data. The error bars give 95\% confidence limits for each median. Each 
datapoint is the outcome of averaging over 75 USA instances and 10 noise 
realizations per USA instance\vspace{0.in}.}
\end{figure}

\begin{figure}[!]
\vspace{0.20in}
\includegraphics[scale=0.325]{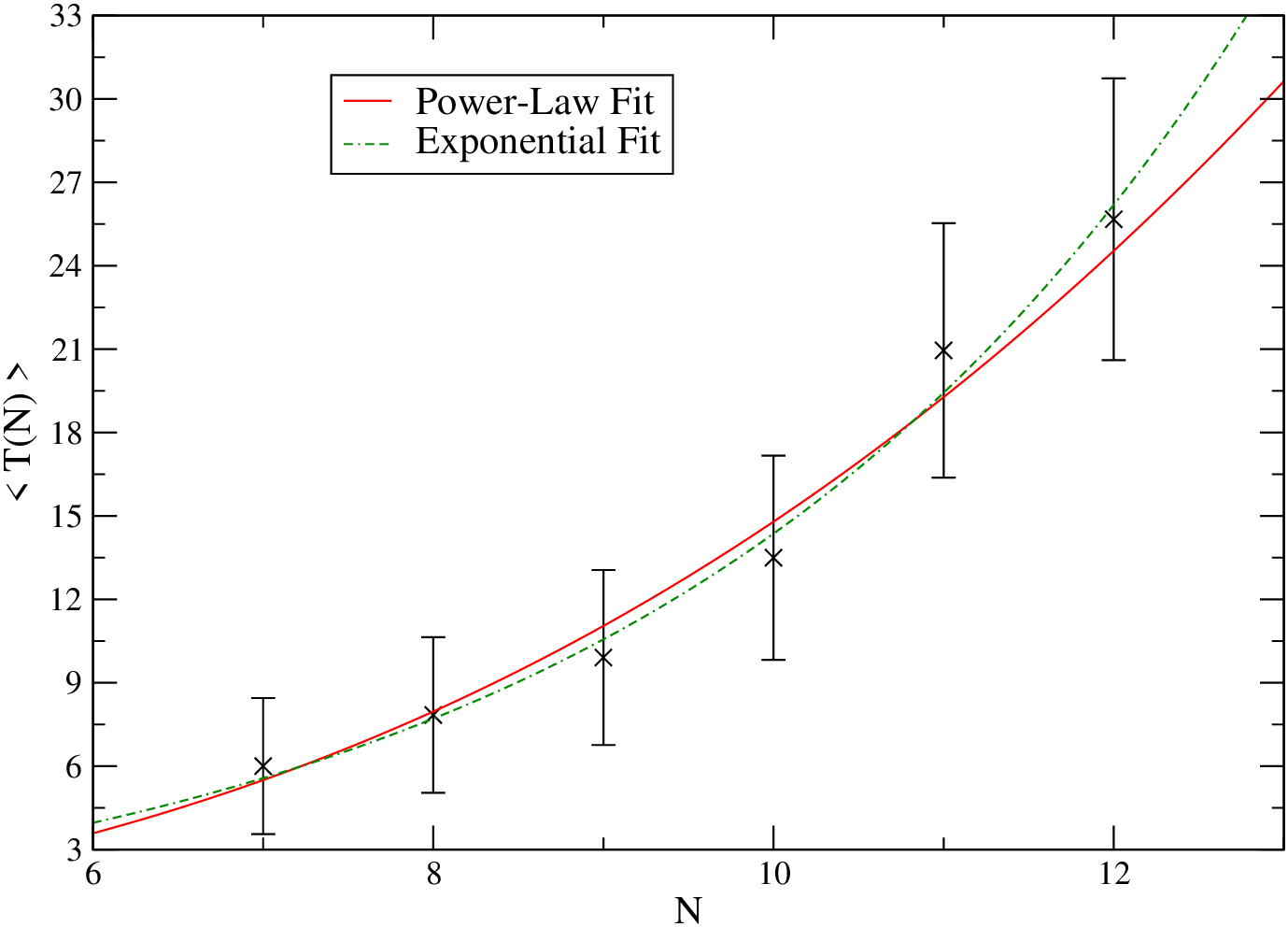}
\vspace{0.1in}
\caption{\label{fig13}Simulation results for QuAdS with noise polarization 
along all 3 directions and mean noise power $\overline{P}=0.005$. Plotted 
are the noise-averaged median runtime $\langle T(N)\rangle$ (dimensionless 
units) versus the number of bits $N$. The solid line is the best power-law 
fit to the data and the dash-dot line is the best exponential fit to the
data. The error bars give 95\% confidence limits for each median. Each 
datapoint is the outcome of averaging over 75 USA instances and 10 noise 
realizations per USA instance\vspace{0.in}.}
\end{figure}

We now present our simulation results. For each mean noise power
$\overline{P}$ we present 4 figures and 2 tables. Each of the figures
contains our numerical results for $\langle T(N)\rangle$ for a particular
noise polarization type (x, y, z, 3), along with two fits to the data:
(i) a power-law fit $\langle T(N)\rangle = aN^{b}$; and (ii) an exponential 
fit $\langle T(N)\rangle = a\left[\,\exp (bN) -1\right]$. Information about 
the best fits to the data are collected in the two tables, one table for each 
of the fitting functions. Each table contains: (i) the best fit parameters; 
(ii) the chi-squared for the fit $\chi_{fit}^{2}$; and (iii) the probability
$P(\chi^{2}>\chi^{2}_{fit})$ that, assuming the fitting function correctly 
describes the scaling of $\langle T(N) \rangle$ versus $N$, a 
sampling of $\langle T(N)\rangle$ would yield a $\chi^{2} > \chi_{fit}^{2}$. 
The closer this probability is to 1, the more consistent the data-set 
is with the fitting function. As such, it provides a quantitative measure
of the quality of fit to the data. Note that, due to the finite-size of 
the data-set, more than one fitting function can be consistent with the data. 
To help the reader locate specific results, we provide the following roadmap 
through the figures and tables.
\begin{enumerate}
\item Figures~\ref{fig2}, \ref{fig3}, \ref{fig4}, and \ref{fig5} correspond to 
x, y, z, and 3-type noise with $\overline{P} = 0.001$, respectively. 
Tables~\ref{table1} and \ref{table2} give the specifics of the power-law
and exponential fits through these data-sets, respectively. 
Table entries are ordered according to increasing scaling-exponent $b$.
\item Figures~\ref{fig6}, \ref{fig7}, \ref{fig8}, and \ref{fig9} correspond to 
x, y, z, and 3-type noise with $\overline{P} = 0.003$, respectively. 
Tables~\ref{table3} and \ref{table4} give the specifics of the power-law
and exponential fits through these data-sets, respectively. 
Table entries are ordered according to increasing scaling-exponent $b$.
\item Figures~\ref{fig10}, \ref{fig11}, \ref{fig12}, and \ref{fig13} 
correspond to x, y, z, and 3-type noise with $\overline{P} = 0.005$, 
respectively. Tables~\ref{table5} and \ref{table6} give the specifics of 
the power-law and exponential fits through these data-sets, respectively. 
Table entries are ordered according to increasing scaling-exponent $b$.
\end{enumerate}
Examination of these figures and tables shows that the power-law and
exponential fits \textit{both\/} provide excellent fits to the numerical
results. One sees that the probability $P(\chi^{2} > \chi^{2}_{fit})$,
and hence the quality of fit, shows a slight decrease with
increasing mean noise power $\overline{P}$, and that the rate of decrease
in the fit quality is largest for y-type noise. One also notes that the quality
of the exponential fit decreases \textit{more slowly} than does the quality of 
the power-law fit. Further discussion of these results is given in 
Section~\ref{sec5}.

\begin{table}[ht]
\caption{\label{table1}Summary of best-fit parameters for power-law scaling 
curves for $\meanP = 0.001$. For comparison, best-fit parameters for 
noiseless QuAdS are also included\vspace{0.10in}.}
\begin{ruledtabular}
\begin{tabular}{c|c|c|c|c|c}
$\meanP$ & noise type & $a$ & $b$ & $\chi^{2}_{fit}$ & 
                                $P(\chi^{2}>\chi^{2}_{fit})$ \\ \hline
0.000 & --- & $1.1966\times 10^{-1}$ & $2.0034$ & $0.092$ & $0.9999844$\\\hline
0.001 & x  & $1.1447\times 10^{-1}$ & $2.0267$ & $0.083$ & $0.9999886$ \\
0.001 & 3  & $1.0635\times 10^{-1}$ & $2.0594$ & $0.064$ & $0.9994915$ \\
0.001 & z  & $1.0395\times 10^{-1}$ & $2.0695$ & $0.063$ & $0.9995085$ \\
0.001 & y  & $9.6766\times 10^{-2}$ & $2.1075$ & $0.078$ & $0.9992617$  
\end{tabular}
\end{ruledtabular}
\end{table} 

\begin{table}[ht]
\caption{\label{table2}Summary of best-fit parameters for exponential scaling 
curves for $\meanP = 0.001$\vspace{0.10in}.}
\begin{ruledtabular}
\begin{tabular}{c|c|c|c|c|c}
$\meanP$ & noise type & $a$ & $b$ & $\chi^{2}_{fit}$ & 
                                $P(\chi^{2}>\chi^{2}_{fit})$ \\ \hline
0.001 & x  & $3.81403$ & $0.142561$ & $0.216$ & $0.9998065$ \\
0.001 & 3  & $2.60771$ & $0.172331$ & $0.075$ & $0.9993111$ \\
0.001 & z  & $2.46414$ & $0.176985$ & $0.059$ & $0.9995798$ \\
0.001 & y  & $2.42677$ & $0.179578$ & $0.085$ & $0.9991179$  
\end{tabular}
\end{ruledtabular}
\end{table} 

\begin{table}[ht]
\caption{\label{table3}Summary of best-fit parameters for power-law scaling 
curves for $\meanP = 0.003$. For comparison, best-fit parameters for 
noiseless QuAdS are also included\vspace{0.10in}.}
\begin{ruledtabular}
\begin{tabular}{c|c|c|c|c|c}
$\meanP$ & noise type & $a$ & $b$ & $\chi^{2}_{fit}$ & 
                                $P(\chi^{2}>\chi^{2}_{fit})$ \\ \hline
0.000 & --- & $1.1966\times 10^{-1}$ & $2.0034$ & $0.092$ & $0.9999844$\\\hline
0.003 & x  & $9.3604\times 10^{-2}$ & $2.1304$ & $0.052$ & $0.9999971$ \\
0.003 & z  & $8.4209\times 10^{-2}$ & $2.1774$ & $0.050$ & $0.9996975$ \\
0.003 & 3  & $7.8825\times 10^{-2}$ & $2.2077$ & $0.083$ & $0.9991666$ \\
0.003 & y  & $2.9821\times 10^{-2}$ & $2.6792$ & $0.570$ & $0.9663310$  
\end{tabular}
\end{ruledtabular}
\end{table} 

\begin{table}[ht]
\caption{\label{table4}Summary of best-fit parameters for exponential scaling 
curves for $\meanP = 0.003$\vspace{0.10in}.}
\begin{ruledtabular}
\begin{tabular}{c|c|c|c|c|c}
$\meanP$ & noise type & $a$ & $b$ & $\chi^{2}_{fit}$ & 
                                $P(\chi^{2}>\chi^{2}_{fit})$ \\ \hline
0.003 & x  & $3.24309$ & $0.157716$ & $0.169$ & $0.9999056$ \\
0.003 & z  & $2.10817$ & $0.193125$ & $0.036$ & $0.9998428$ \\
0.003 & 3  & $2.05131$ & $0.195818$ & $0.075$ & $0.9993165$ \\
0.003 & y  & $0.847514$ & $0.285049$ & $0.283$ & $0.9908794$  
\end{tabular}
\end{ruledtabular}
\end{table} 

\begin{table}[ht]
\caption{\label{table5}Summary of best-fit parameters for power-law scaling 
curves for $\meanP = 0.005$. For comparison, best-fit parameters for 
noiseless QuAdS are also included\vspace{0.10in}.}
\begin{ruledtabular}
\begin{tabular}{c|c|c|c|c|c}
$\meanP$ & noise type & $a$ & $b$ & $\chi^{2}_{fit}$ & 
                                $P(\chi^{2}>\chi^{2}_{fit})$ \\ \hline
0.000 & --- & $1.1966\times 10^{-1}$ & $2.0034$ & $0.092$ & $0.9999844$\\\hline
0.005 & x  & $6.679\times 10^{-2}$ & $2.2901$ & $0.132$ & $0.9996794$ \\
0.005 & z  & $3.5862\times 10^{-2}$ & $2.5964$ & $0.389$ & $0.9833500$ \\
0.005 & 3  & $2.4823\times 10^{-2}$ & $2.7752$ & $0.484$ & $0.9750332$ \\
0.005 & y  & $6.6069\times 10^{-3}$ & $3.429$ & $1.052$ & $0.9017628$  
\end{tabular}
\end{ruledtabular}
\end{table} 

\begin{table}[ht]
\caption{\label{table6}Summary of best-fit parameters for exponential scaling 
curves for $\meanP = 0.005$\vspace{0.10in}.}
\begin{ruledtabular}
\begin{tabular}{c|c|c|c|c|c}
$\meanP$ & noise type & $a$ & $b$ & $\chi^{2}_{fit}$ & 
                                $P(\chi^{2}>\chi^{2}_{fit})$ \\ \hline
0.005 & x  & $2.25018$ & $0.189744$ & $0.178$ & $0.9993283$ \\
0.005 & z  & $1.12768$ & $0.258381$ & $0.256$ & $0.9924477$ \\
0.005 & 3  & $0.859322$ & $0.287436$ & $0.253$ & $0.9926288$ \\
0.005 & y  & $0.498037$ & $0.357153$ & $0.707$ & $0.9504644$  
\end{tabular}
\end{ruledtabular}
\end{table} 

\section{\label{sec5} Discussion}

\subsection{\label{sec5a} Summary}

We examine here what can be learned from the results presented in
Section~\ref{sec4}. To begin, we note that both power-law scaling 
$\langle T(N) \rangle = aN^{b}$ and 
exponential scaling $\langle T(N) \rangle = a\left[\exp (bN) - 1\right]$ 
provide excellent fits to our simulation results, although the fit
parameters are noise dependent. Tables~\ref{table1}--\ref{table6} show that 
the scaling exponent $b$ increases with increasing mean noise power $\meanP$, 
and the rate at which it increases depends on the noise polarization, or 
direction along which the noise field $\bNi (t)$ fluctuates. Clearly, as the 
scaling exponent increases, QuAdS performance decreases non-linearly. We 
simulated noise with noise polarization: (i) along $\hat{\mathbf{x}}$; 
(ii) along $\hat{\mathbf{y}}$; (iii) along $\hat{\mathbf{z}}$; and (iv) along 
all 3 directions simultaneously. At the noise power levels considered, x-type 
noise was found to have the smallest impact on QuAdS performance, while y-type 
noise was found to cause the largest slowdown. As noted in Section~\ref{sec4}, 
although the quality of both the power-law and exponential fits is excellent 
in all cases considered, we did see a slight reduction in the quality of fit 
as noise power increased for all noise-types except x-type noise. 
We also noted (see Tables~\ref{table5} and \ref{table6} especially) that the
quality of the exponential fit decreased at a slightly slower rate than did
the power-law fit. It would clearly be of interest to extend the simulations 
to larger noise power levels to examine: (i) how rapidly the scaling exponent 
$b$ increases with noise power; and (ii) how quickly the quality of fit 
for both power-law and exponential fits deteriorates with increasing noise 
power. Given that a quantum computer will eventually crossover to classical
behavior with sufficient noise power, and should classical algorithms truly
require exponential time on the randomly generated hard instances considered
in Ref.~\cite{fa1} and here, then one would anticipate a power-law fit to
eventually become inconsistent with the simulation data at sufficient noise
power and the exponential fit to provide the better fit. Beyond this 
threshold value of noise power, QuAdS would then be expected to have
performance comparable with classical algorithms. Observation of such a 
crossover would be extremely important as it would give a direct measure of 
how much noise QuAdS can tolerate. In the following subsection we show that
noise-induced dephasing leads to decoherence in QuAdS, estimate how much 
decoherence is present in our simulations, and derive an upper bound for 
the noise-averaged QuAdS success probability. 

\subsection{\label{sec5b} Noise-Induced Decoherence in QuAdS}

Although our focus in this paper is on QuAdS, the following analysis can be 
adapted to the more general situation of noisy quantum adiabatic dynamics. 
We will report on this elsewhere.

The total Hamiltonian for noisy QuAdS is (see eq.~(\ref{totHam})),
\begin{displaymath}
\mathcal{H}(t) = H(t) + H_{int}(t) \hspace{0.1in} ,
\end{displaymath}
with $H(t)$ and $H_{int}(t)$ given in eqs.~(\ref{timeHam}) and (\ref{Hint}), 
respectively.  In the Schrodinger picture, the dynamics is driven by
\begin{displaymath}
i\frac{\partial}{\partial t} |\psi\rangle = \mathcal{H}(t)|\psi\rangle
  \hspace{0.1in} .
\end{displaymath}
Transforming to an interaction-like picture (the time-ordering symbol is
suppressed),
\begin{equation}
\label{intpic}
|\psi (t)\rangle = \exp \left[\, -i\int_{0}^{t}\, ds\, H_{int}(s)\,\right]\,
                    |\chi (t)\rangle \hspace{0.1in} ,
\end{equation}
the equation of motion for $|\chi (t)\rangle$ is found to be
\begin{equation}
\label{chischeq}
i\frac{\partial}{\partial t} |\chi \rangle = \overline{H}(t)|\chi\rangle
                           \hspace{0.1in},
\end{equation} 
where
\begin{equation}
\label{intpicHam}
\overline{H}(t) = e^{ i\int_{0}^{t} ds H_{int}(s)}\, H(t)\, e^{ -i\int_{0}^{t} 
                      ds H_{int}(s)} \hspace{0.1in} .
\end{equation}
Using the well-known identity
\begin{equation}
\label{opident}
e^{\xi A}\, B\,e^{-\xi A} =
  B + \xi\left[ A,B\right] +\frac{\xi^{2}}{2}\left[ A, \left[ A,B\right]
                 \right] + \cdots 
\end{equation}
with $B = H(t)$ and
\begin{eqnarray}
\label{Aop}
\xi A & = & i\int_{0}^{t}\, ds\, H_{int}(s) \nonumber \\
      & = & \left[\, -i\gamma\,\int_{0}^{t}\, x(s)\, ds\,\right]
             \sum_{i=1}^{N}\,\sigma_{x}^{i} \hspace{0.1in} ,
\end{eqnarray}
identifies
\begin{equation}
\label{xidef}
\xi = -i\gamma\,\int_{0}^{t}\, x(s)\, ds \hspace{0.1in} ,
\end{equation}
and
\begin{equation}
\label{Adef}
A = \sum_{i=1}^{N}\,\sigma_{x}^{i} \hspace{0.1in} .
\end{equation}
To make our discussion concrete, I will assume the presence of  x-noise: 
$\mathbf{N}_{i} (t) = x(t)\hat{\mathbf{x}}$; and set $\gamma_{i} = \gamma$. 
Combining eqs.~(\ref{intpicHam}) 
through (\ref{Adef}) gives
\begin{equation}
\label{pertexp}
\overline{H}(t) = H(t) - \left( i\gamma\int_{0}^{t} x(s)ds\right)
                    \left[\sum_{i=1}^{N}\,\sigma_{x}^{i}, H(t)\right]
                      + \cdots .
\end{equation}
In the simplest approximation, only the first term in eq.~(\ref{pertexp})
is kept. This corresponds to the limit of weak noise which is appropriate
for our simulations. Clearly eq.~(\ref{pertexp}) lends itself to a systematic
evaluation of corrections to the weak noise approximation. Evaluating
eq.~(\ref{intpic}) at the end of the quantum search $t=T$ gives
\begin{equation}
\label{fnlwavefcn}
|\psi (T)\rangle = \exp\left[ -i\int_{0}^{T} ds\, H_{int}(s)\right]
                    |\chi (T)\rangle \hspace{0.1in} ,
\end{equation} 
and in the weak noise approximation, eq.~(\ref{chischeq}) becomes
\begin{equation}
\label{fnlchischeq}
i\frac{\partial}{\partial t} |\chi\rangle = H(t)|\chi\rangle \hspace{0.1in} .
\end{equation}
Thus, in the weak noise limit, the time-development of $|\chi (t)\rangle$ is 
driven by the Hamiltonian $H(t)$ for noiseless QuAdS.

To evaluate eq.~(\ref{fnlwavefcn}), we introduce the instantaneous
eigenvalues and eigenstates of $H_{int}(t)$:
\begin{equation}
\label{Hintvalstates}
H_{int}(t)|E_{\mbox{\boldmath $\scriptstyle\sigma$}}(t)\rangle = 
              E_{\mbox{\boldmath $\scriptstyle\sigma$}}(t) 
               |E_{\mbox{\boldmath $\scriptstyle\sigma$}}
                  (t)\rangle \hspace{0.1in} .
\end{equation}
Following the usual decoherence language \cite{zur}, we refer to the
$|E_{\mbox{\boldmath $\scriptstyle\sigma$}}(t)\rangle$ as the pointer basis 
states. For x-noise, the pointer basis states are time-independent:
\begin{eqnarray}
\label{pointbas}
|E_{\mbox{\boldmath $\scriptstyle\sigma$}}(t)\rangle & = & 
        |E_{\mbox{\boldmath $\scriptstyle\sigma$}}\rangle \nonumber \\ 
  & = & |\sigma_{1}\cdots\sigma_{N}\rangle_{x} \nonumber \\
  & = & |\sigma_{1}\rangle_{x}\otimes\cdots\otimes |\sigma_{N}\rangle_{x}
        \hspace{0.1in} ,
\end{eqnarray}
where $\sigma_{i} = \pm 1$ and $\mbox{\boldmath $\sigma$} = (\sigma_{1},
\ldots ,\sigma_{N})$. It follows from our expression for $H_{int}(t)$
(eqs.~(\ref{Hint}), (\ref{Hintvalstates}), and (\ref{pointbas})) that
\begin{equation}
\label{eigenval}
E_{\mbox{\boldmath $\scriptstyle\sigma$}}(t) = -\gamma x(t)\sigma 
\hspace{0.1in} ,
\end{equation}
with  
\begin{equation}
\label{sigeig}
\sigma = \sum_{i=1}^{N}\sigma_{i} \hspace{0.1in} .
\end{equation}
Dividing the time interval $[0, T]$ into $N$ subintervals of duration
$\epsilon = T/N$, and noting that $[H_{int}(t),\; H_{int}(t^{\prime})] = 0$
for x-noise allows us to write the exponential in eq.~(\ref{fnlwavefcn}) as
\begin{equation}
\label{factorexp}
\exp\left[ -i\int_{0}^{T} ds H_{int}(s)\right] = \prod_{i=1}^{N}U(t_{i},
                                                   t_{i-1})
\hspace{0.1in} ,
\end{equation}
where
\begin{equation}
\label{factorelem}
U(t_{i},t_{i-1}) = \exp\left[ -i\int_{t_{i-1}}^{t_{i}} ds H_{int}(s)\right]
  \hspace{0.1in} .
\end{equation}
Using eq.~(\ref{factorexp}) and inserting the completeness relation for the 
pointer basis states $\{ |E_{\mbox{\boldmath $\scriptstyle\sigma$}}\rangle\}$ 
into eq.~(\ref{fnlwavefcn}) gives
\begin{eqnarray}
\label{pointbaswavefcn}
\lefteqn{\hspace{-0.25in}|\psi (T)\rangle  =  } \nonumber \\
          & & \hspace{-0.25in}         \sum_{\sigma_{1}\cdots\sigma_{N}}\, 
                        |\sigma_{1}\cdots\sigma_{N}\rangle
                        \langle\sigma_{1}\cdots\sigma_{N}|
                        \prod_{i=1}^{N}
                          U(t_{i},t_{i-1})|\chi (T)\rangle .
\end{eqnarray}
Using eqs.~(\ref{factorelem}) and (\ref{Hintvalstates})  gives
\begin{eqnarray}
\label{tempres}
\lefteqn{\hspace{-0.25in}\langle \sigma_{1}\cdots\sigma_{N}|
               U(t_{i},t_{i-1}) = } \nonumber \\
 & & \hspace{0.45in}\langle \sigma_{1}\cdots\sigma_{N}| 
\exp\left[ -i\int_{t_{i-1}}^{t_{i}}
                         ds E_{\mbox{\boldmath $\scriptscriptstyle\sigma$}}
                     (s)\right] .
\end{eqnarray}
Finally, using eq.~(\ref{tempres}) in eq.~(\ref{pointbaswavefcn}) gives
\begin{equation}
|\psi (T)\rangle  = 
                    \sum_{\sigma_{1}\cdots\sigma_{N}}\, 
                        |\sigma_{1}\cdots\sigma_{N}\rangle
           \exp\left[ -i\phi_{\mbox{\boldmath $\scriptscriptstyle\sigma$}}(T)
             \right]           
           a_{\mbox{\boldmath $\scriptstyle\sigma$}}(T) 
                       \hspace{0.1in} ,
\end{equation}
where
\begin{equation}
\label{stocphaseshft}
\phi_{\mbox{\boldmath $\scriptstyle\sigma$}}(T) = \int_{0}^{T}ds 
   E_{\mbox{\boldmath $\scriptstyle\sigma$}}(s) = -\gamma \sigma\int_{0}^{T}
    x(s)ds 
   \hspace{0.1in} ,
\end{equation}
and 
\begin{equation}
a_{\mbox{\boldmath $\scriptstyle\sigma$}}(T) = \langle\sigma_{1}\cdots
\sigma_{N}|
                                        \chi (T)\rangle \hspace{0.1in} .
\end{equation}
The final density matrix $\rho (T) = |\psi (T)\rangle\langle\psi (T)|$ has
matrix elements (in the pointer basis)
\begin{eqnarray}
\rho_{\mbox{\boldmath $\scriptstyle\sigma$},
\mbox{\boldmath $\scriptstyle\sigma$}^{\prime}}
    (T) & = &  \exp\left[ -i\bigGama \right]
               a_{\mbox{\boldmath $\scriptstyle\sigma$}}(T) 
                a^{\ast}_{\mbox{\boldmath $\scriptstyle\sigma$}^{\prime}}(T) ,
\end{eqnarray}
where
\begin{eqnarray}
\label{Gamdef}
\bigGama & = & 
           \phi_{\mbox{\boldmath $\scriptstyle\sigma$}}(T)
        -  \phi_{\mbox{\boldmath $\scriptstyle\sigma^{\prime}$}}(T)
             \nonumber \\
  & = & -\gamma\left(\sigma - \sigma^{\prime}\right)\int_{0}^{T} x(s) ds
    \hspace{0.1in} .
\end{eqnarray}
As noted above, in the weak noise approximation, eq.~(\ref{fnlchischeq}) 
indicates that $a_{\mbox{\boldmath $\scriptstyle\sigma$}}(T)$ is determined by 
the noiseless QuAdS dynamics. Thus if $|\psi (0)\rangle = |\chi (0)\rangle$ is 
initially equal to the initial groundstate $|E_{g}(0)\rangle$, then 
$|\chi (T)\rangle = |E_{g}(T)\rangle$ and so $a_{\mbox{\boldmath 
$\scriptstyle\sigma$}}(T) = \langle \sigma_{1}\cdots\sigma_{N}|E_{g}(T) 
\rangle$. Thus all noise dependence in this approximation appears in the 
stochastic phases $\{\phi_{\mbox{\boldmath $\scriptstyle\sigma$}}(T)\}$ which 
clearly depend on $\mbox{\boldmath $\sigma$}$ (see eq.~(\ref{stocphaseshft})). 
The stochastic character of the noise requires us to represent
our quantum system by an ensemble in which each element of the ensemble is
our quantum system in the presence of a particular noise realization. As we
do not know which element of the ensemble will correspond to our quantum
system on a given run of QuAdS, we must average over the ensemble to
determine the expected performance of QuAdS in the presence of noise. 
We now show that these stochastic phases lead to a 
suppression of the off-diagonal matrix elements ($\mbox{\boldmath $\sigma$} 
\neq \mbox{\boldmath $\sigma^{\prime}$}$) of $\rho (T)$ when the 
noise-average is carried out. The noise-averaged density matrix 
$\overline{\rho}_{\mbox{\boldmath $\scriptstyle\sigma$},
\mbox{\boldmath $\scriptstyle\sigma$}^{\prime}}(T)$ is thus
\begin{equation}
\label{noyzavrho}
\overline{\rho}_{\mbox{\boldmath $\scriptstyle\sigma$},
\mbox{\boldmath $\scriptstyle\sigma$}^{\prime}}(T) = \decofac \, 
a_{\mbox{\boldmath $\scriptstyle\sigma$}}(T)
                 a^{\ast}_{\mbox{\boldmath $\scriptstyle\sigma$}^{\prime}}(T) 
  \hspace{0.1in} ,
\end{equation}
where
\begin{equation}
\decofac = \overline{\exp\left[ -i\bigGama\right]}
\end{equation}
is the decoherence factor. In the adiabatic limit, the thermal relaxation time 
(a.k.a.~noise correlation time) $\tau$ satisfies $\tau\ll T$. We divide the 
integration interval $[0,\: T]$ appearing in $\bigGama$ (eq.~(\ref{Gamdef}))
into $M=T/\tau$ subintervals of duration $\tau$. This renders $\bigGama$ into 
a sum of uncorrelated random variables $\bigGama (j)$:
\begin{equation}
\label{gamdef}
\bigGama = \sum_{j=1}^{M}\,\bigGama (j) \hspace{0.1in} ,
\end{equation}
where
\begin{equation}
\label{gamideftn}
\bigGama (j) = -\gamma\left(\sigma - \sigma^{\prime}\right) 
                 \int_{(j-1)\tau}^{j\tau}x(s)ds \hspace{0.1in} .
\end{equation}
Since the noise is stationary, the set of $\{\bigGama (j)\}$ have identical
probability distributions. If the $\{\bigGama (j)\}$ are not only 
uncorrelated, but also statistically independent, it follows from the
central limit theorem that $\bigGama$ will have a Gaussian probability
distribution with mean $\overline{\bigGama}$ and variance 
$\overline{\bigGama^{2}}$. From eqs.~(\ref{gamdef}) and (\ref{gamideftn}),
and the fact that the noise $x(t)$ has zero mean $\overline{x(t)} = 0$,
it follows that $\overline{\bigGama}=0$. From eq.~(\ref{Gamdef}),
\begin{equation}
\bigGama^{2} = \gamma^{2}\left(\sigma - \sigma^{\prime}\right)^{2}
                  \int_{0}^{T}ds\int_{0}^{T}ds^{\prime} x(s)x(s^{\prime})
  \hspace{0.1in} .
\end{equation}
Averaging over the noise gives
\begin{equation}
\label{gamsvar}
\overline{\bigGama^{2}} = \gamma^{2}\left(\sigma - \sigma^{\prime}\right)^{2}
                             \int_{0}^{T}ds\int_{0}^{T}ds^{\prime}
                              \overline{x(s)x(s^{\prime})} \hspace{0.1in} .
\end{equation}
For our noise model, the noise correlation function 
$\overline{x(s)x(s^{\prime})}$ is (see Section~\ref{sec3a}):
\begin{equation}
\label{noyzcorfcn}
\overline{x(s)x(s^{\prime})} = \sigma_{x}^{2}\, h(s-s^{\prime}) 
  \hspace{0.1in} .
\end{equation}
To avoid confusion with $\sigma$ defined in eq.~(\ref{sigeig}), we have 
written $\sigma_{x}^{2}$ for the variance of x-noise (denoted by 
$\sigma^{2}$ in  Sections~\ref{sec3} and \ref{sec4}); and recall that 
$h(s-s^{\prime})$ is our 
square pulse noise fluctuation profile of unit height and width $2\tau$. 
Using eq.~(\ref{noyzcorfcn}) in eq.~(\ref{gamsvar}) and carrying out the 
integrations gives:
\begin{equation}
\label{explicitvar}
\overline{\bigGama^{2}} = 4\tau T\left(\sigma -\sigma^{\prime}\right)^{2}
                             \gamma^{2}\sigma_{x}^{2} \hspace{0.1in} .
\end{equation}
We now have the ingredients needed to evaluate the decoherence factor 
$\decofac$:
\begin{eqnarray}
\decofac  & = & \int_{-\infty}^{\infty}
         \frac{d\bigGama}{\sqrt{2\pi\overline{\bigGama^{2}}}}\,
          \exp\left[ -\frac{\bigGama^{2}}{2\overline{\bigGama^{2}}}
           \right]\,\exp\left[ -i\bigGama\right] \nonumber \\
 & = & \exp\left[ - \frac{\overline{\bigGama^{2}}}{2}\right] 
         \hspace{0.1in} .
\end{eqnarray}
Using eq.~(\ref{explicitvar}) gives
\begin{equation}
\label{decoformula}
\decofac  = \exp\left[ -2\gamma^{2}\tau T\sigma_{x}^{2}\left(\sigma - 
      \sigma^{\prime}\right)^{2}\right] \hspace{0.1in} .
\end{equation}
A number of remarks are in order.\\ 

\noindent (1) For quantum adiabatic dynamics, $T\gg 1$. Thus for non-zero
$\sigma_{x}^{2}$, $\tau$, and $\gamma$, the decoherence factor $\decofac$ is
exponentially small. Thus a quantum state that is a superposition of pointer
basis states will undergo an effective wavefunction collapse and $\rho (T)$ 
will become effectively diagonal in the pointer basis. Thus we get the 
standard decoherence phenomenology \cite{zur} from our noise model. As we 
have seen, the noisy background field $\mathbf{N}_{i}(t)$ generates a 
stochastic phase $\phi_{\mbox{\boldmath $\scriptstyle\sigma$}}(T)$ which is 
different for the different pointer basis states $|E_{\mbox{\boldmath 
$\scriptstyle\sigma$}}(t) \rangle$, and which causes suppression of the 
off-diagonal matrix elements of $\rho (T)$ when we average over the noise. As 
noted above, since one doesn't know what noise realization will appear on any 
particular run of QuAdS, one must noise-average to determine the impact of
noise on QuAdS performance. It is worth 
noting that, although the original decoherence analysis \cite{zur} considers
an environment which is a quantum system, it has been conventional wisdom
that noise-induced dephasing can generate the usual decoherence phenomenology
\cite{zur_rmp}, although we are not aware of any explicit demonstration of
this expectation prior to the one presented above. \\

\noindent (2) For QuAdS, the above analysis is relevant whenever $[H_{int}(t),
H(t)]\neq 0$, since then the pointer basis states differ from the
instantaneous eigenstates of $H(t)$. Since the instantaneous eigenstates
must then be a superposition of the pointer basis states, they are vulnerable
to the effective wavefunction collapse produced by the noise-induced
decoherence discussed above. Since such an effective collapse
signals the effective loss of quantum coherence, one would expect
that this decoherence would jeopardize any advantage QuAdS might
acquire due to the quantum nature of its dynamics. It was this naive
expectation that motivated us to do a careful study of the effects of noise
of QuAdS performance. To test this idea, we decided to study noise which 
coupled to the qubits via the Zeeman interaction (see eqn.~(\ref{Hint})), 
and whose polarization was constant: $\hat{\mathbf{N}}_{i}(t) = 
\hat{\mathbf{x}}; \: \hat{\mathbf{y}};\: \hat{\mathbf{z}}$. The pointer basis 
states are then, respectively, the eigenstates of direct products of the 
qubit Pauli operators: $\prod_{i=1}^{N}\otimes\,\sigma_{x}^{i}$; 
$\prod_{i=1}^{N}\otimes\, \sigma_{y}^{i}$; and $\prod_{i=1}^{N}\otimes\,
\sigma_{z}^{i}$. We also recognized  that such an interaction would also 
allow a study of what would happen if the coupling interaction generated a 
pointer basis that varied randomly with time. This case corresponds to our 
3-noise simulation in which the noise fluctuates along all 3 directions 
simultaneously. Refs.~\cite{LO1} and \cite{LO2} 
show that a quantum phase transition occurs during QuAdS near $t/T \sim 0.7$. 
Before (after) this phase transition the quantum dynamics is essentially 
driven by $H_{i}$ ($H_{p}$). As discussed in Section~\ref{sec2b}, the 
eigenstates of $H_{i}$ ($H_{p}$) correspond to qubit spins aligned along 
$\hat{\mathbf{x}}$ ($\hat{\mathbf{z}}$). 
\begin{itemize}
\item For noise with fluctuations along $\bfxh$ and for
$t/T < 0.7$, $[H(t),\: H_{int}(t)] \approx 0$ and the instantaneous energy
eigenstates are essentially the same as the pointer basis states throughout
the first $70\%$ of the quantum evolution. Noise-induced dephasing is thus
only effective during the final $30\%$ of the evolution where $[H(t),\: 
H_{int}(t)]\neq 0$ and the pointer basis differs from the instantaneous
energy eigenstates. Only during the final $30\%$ of the quantum evolution
will noise act to dephase entanglement. 
\item For noise fluctuations along 
$\bfyh$, $[H(t),\: H_{int}(t)] \neq 0$ for all $t$, and the instantaneous 
energy eigenstates \textit{never} correspond to the pointer basis. Thus 
noise-induced dephasing occurs throughout the \textit{entire} dynamical 
evolution. 
\item For z-type noise, $[H(t),\: H_{int}(t)] \approx 0$ during 
the final $30\%$ of the dynamical evolution, while noise-induced dephasing 
occurs over the initial $70\%$ of the evolution. Thus dephasing of 
entanglement occurs over nearly 3/4's of the quantum evolution. 
\end{itemize}
These remarks suggest that z-type noise should have a larger impact on QuAdS 
than x-type noise, and that y-type noise should have the most severe impact of 
the three. As 3-type noise has fluctuations along all 3 directions, it is 
unlikely that noise fluctuations of this type will remain aligned with any one 
of the coordinate axes throughout the entire adiabatic evolution. Thus 3-type 
noise will sample the most damaging y-type fluctuations less often than y-type 
noise, and the more benign x-type fluctuations less often than x-type noise. 
As a result, one would expect 3-type noise to have a less severe impact 
on QuAdS than y-type noise, and more of an impact than x-type noise. This is
what we see in our simulations (see Tables~\ref{table1}--\ref{table6}).
We see that knowledge of the occurrence of a quantum phase transition during 
QuAdS near $t/T \sim 0.7$, together with knowledge of how phase decoherence 
degrades pointer basis superpositions, allows us to understand most of the 
trends in our numerical results. Applying these arguments to a comparison of 
3-type noise with z-type noise does not lead to any simple conclusion as far 
as we can tell. Our simulation results suggest that this case is in fact less 
straightforward as 3-type noise is less damaging to QuAdS performance than 
z-type noise at the lowest power levels simulated, but becomes more 
detrimental to QuAdS at the larger power levels. \\ 

\noindent (3) Using eq.~(\ref{decoformula}), we can estimate the
degree of decoherence in our simulations. From this formula we see that the
least amount of decoherence will occur for $\sigma - \sigma^{\prime} = 1$
(i.~e.~$\decofac$ is largest): 
\begin{equation}
\label{speclcase}
D_{max}^{x} = \exp\left[\, -2\gamma^{2}\tau T\sigma_{x}^{2}\,\right] 
   \hspace{0.1in} .
\end{equation}
For our simulations, $\sigma_{x} = 0.2$, $\tau = 1$, and $\gamma = 1$.
For x-noise with $\overline{P} = 0.005$ and $N = 13$, $<T(13)>\sim 22$ and
$[H_{int}(t),H(t)]\neq 0$ for $0.7T \leq t \leq T$, or $30\%$ of the time
for QuAdS to be carried out. Thus only $0.3T$ should be used in 
eq.~(\ref{speclcase}) since only for this duration did the instantaneous
eigenstates of $H(t)$ differ from the pointer basis. Inserting these values
into eq.~(\ref{speclcase}) gives $D_{max}^{x}\sim 0.590$. Thus for our 
simulation of x-noise, QuAdS still possesses a substantial amount of quantum
coherence. Repeating the above analysis for y-noise we again find
eq.~(\ref{decoformula}) with $\sigma_{x}^{2}\rightarrow\sigma_{y}^{2}$.
Setting ($\sigma -\sigma^{\prime} = 1$), we find:
\begin{equation}
\label{speclycase}
D_{max}^{y} = \exp\left[\, -2\gamma^{2}\tau T\sigma_{y}^{2}\,\right] 
   \hspace{0.1in} .
\end{equation}
For $\overline{P}=0.005$, $<T(12)>\sim 35$, and $[H_{int}(t),H(t)]\neq 0$
throughout all of the QuAdS. Thus, inserting $T\sim 35$ into 
eq.~(\ref{speclycase}), along with $\sigma_{y}=0.2$, and the above values for 
$\tau$ and $\gamma$ gives $D_{max}^{y} \sim 0.06$. Thus y-noise at 
$\overline{P}= 0.005$ produces non-trivial decoherence effects. \\

\noindent (4) Finally, we show that for weak noise $D_{max}$ provides an
upper bound for the noise-averaged success probability of QuAdS. \\

\noindent As explained in Section~\ref{sec2b}, for a USA instance of EC3, the 
groundstate $|E_{g}(T)\rangle$ of the final noiseless QuAdS Hamiltonian 
$H(T)$ (eq.~(\ref{timeHam})) encodes the unique solution to this instance. 
The probability $P_{suc}$ for QuAdS to succeed on this instance is:
\begin{equation}
\label{noiselesssucprob}
P_{suc} = \mathrm{Tr}\, P_{g}\, \rho (T) \hspace{0.1in} ,
\end{equation}
where $P_{g} = |E_{g}(T)\rangle\langle E_{g}(T)|$ is the projection operator
onto the final groundstate $|E_{g}(T)\rangle$ and $\rho (T) = |\psi (T)\rangle
\langle\psi (T)|$ is the final density matrix. We can assess the impact of 
noise on the performance of QuAdS by evaluating the noise-averaged
success probability $\overline{P_{suc}}$:
\begin{equation}
\overline{P_{suc}}\hspace{0.1in} = \hspace{0.1in}\overline{\mathrm{Tr}P_{g}
\rho (T)} \hspace{0.1in} .
\end{equation}
Since $P_{g}$ projects onto the final groundstate of the noiseless QuAdS
Hamiltonian, it does not depend on noise and so:
\begin{eqnarray}
\lefteqn{\hspace{-0.335in}\overline{P_{suc}}  =  \mathrm{Tr}\, P_{g}\,
   \overline{\rho}(T)} \nonumber  \\
 & & \hspace{-0.2in}  =  \sum_{\bfsig ,\bfsig^{\prime}}\left( a_{\bfsig}(T)
         a^{\ast}_{\bfsig^{\prime}}(T)\right)\,
          \left\{ D_{\bfsig^{\prime},\bfsig}\, a_{\bfsig^{\prime}}(T)
           a^{\ast}_{\bfsig}(T)\right\} ,
\end{eqnarray}
where we: (i) have carried out the trace using the pointer basis states
(eqs.~(\ref{Hintvalstates}) and (\ref{pointbas})); and (ii) used 
eq.~(\ref{noyzavrho}) and the definition of $a_{\bfsig}(T)$ given below 
eq.~(\ref{Gamdef}). Noting that the set of $\{ a_{\bfsig}(T)
a^{\ast}_{\bfsig^{\prime}}(T)\}$ are the matrix elements of the density
matrix $\rho_{g}(T) =|E_{g}(T)\rangle\langle E_{g}(T)|$ in the pointer basis
representation, we have
\begin{eqnarray}
\label{finlnoyzsucprob}
\overline{P_{suc}} & = & \sum_{\bfsig ,\bfsig^{\prime}}\,
                          D_{\bfsig^{\prime},\bfsig}\,
                         (\rho_{g}(T))_{\bfsig ,\bfsig^{\prime}}\,
                          (\rho_{g}(T))_{\bfsig^{\prime},\bfsig} \nonumber \\
  & \leq & D_{max}\mathrm{Tr}\left(\rho_{g}(T)\right)^{2} \nonumber \\
  & \leq & D_{max} \hspace{0.1in} .
\end{eqnarray}
Here we have used that $\rho_{g}(T)$ describes a pure state so that 
$(\rho_{g}(T))^{2}= \rho_{g}(T)$. From eqs.~(\ref{speclcase}) and 
(\ref{speclycase}) we see that the noise-averaged success probability for 
QuAdS is exponentially sensitive to the noise parameters appearing in 
$D_{max}$. It is clearly of interest to extend the above analysis beyond the 
limit of weak noise. We leave this extension to future work.

\subsection{\label{sec5c} Future Work}

A number of directions for future work suggest themselves. As mentioned above,
we would like to extend our simulations to larger noise power levels to 
examine: (i)~how rapidly the scaling exponent $b$ increases with noise power; 
(ii)~the degree to which the quality of the power-law and exponential fits 
continues to decrease with noise power; and (iii)~how the quality of the
exponential-fit compares with that of the power-law fit as noise power is 
increased. The aim of this latter point being to see if we can observe
whether one of the two types of scaling laws begins to provide a significantly 
better fit than the other at some higher noise power level. We would also like 
to examine quantitatively how QuAdS performance is affected by noise which 
varies from one qubit-site to another. One would expect that, for this type of 
noise, noise-induced decoherence would be more effective at hampering QuAdS 
performance than the uniform noise we have considered in this paper. It would
also be interesting to extend the weak noise analysis of the upper bound on 
the noise-averaged success probability for QuAdS to stronger noise. Finally, 
Farhi et.\ al.\ \cite{fa3} have argued that varying the path from $H_{i}$ to 
$H_{p}$ might improve QuAdS performance. Noise can be thought to implement a 
random path variation, and so it might be interesting to examine whether a 
noise parameter regime exists where the performance enhancement of 
Ref.~\cite{fa3} might occur. As pointed out in the Introduction, we did find
noise realizations which did reduce the runtime of QuAdS on a given USA
instance, but it appears that the predominant effect of noise at the power 
levels that we considered is to slow down QuAdS. It might be worthwhile to 
examine whether noise-improved QuAdS might occur at small noise power levels.

\section{\label{sec6} Summary}

In this paper we have presented the results of a large-scale simulation
of QuAdS in the presence of noise. We determined the noise-averaged
median runtime $\langle T(N)\rangle$ for QuAdS to succeed in solving USA
instances of the NP-Complete problem N-Bit Exact Cover 3. Clear evidence was 
found of the algorithm's sensitivity to noise. We simulated noise with 4 
different types of polarization, and our results are the outcome of 
approximately 300,000 integrations of the Schrodinger equation. The scaling 
relation for $\langle T(N)\rangle$ versus the number of bits $N$ was fit with 
both a power-law scaling $\langle T(N)\rangle = aN^{b}$ and an exponential 
scaling $\langle T(N) \rangle = a\left[\exp (bN) -1\right]$. \textit{Both\/}
scaling relations provided excellent fits to the simulation results, although
the quality of the fits were found to decrease slightly with increasing noise 
power. The quality of the exponential fit decreased at a slightly slower
rate than did that of the power-law fit. The variation of the scaling 
parameters $a$ and $b$ with mean noise power and noise polarization was 
determined. Our simulation results are summarized in 
Tables~\ref{table1}--\ref{table6}. These tables order the noise types 
according to which type most slowed down QuAdS. We also showed how
noise-induced dephasing can cause decoherence in the dynamics of QuAdS, 
estimate the amount of decoherence present in our simulations of nosiy
QuAdS, and derive an upper bound for the noise-averaged QuAdS success
probability in the weak noise limit that is appropriate for our simulations.

\begin{acknowledgments}
I would like to thank T. Howell III for continued support, the National 
Science Foundation for support provided through Grant No.\ NSF-PHY-0112335, 
and the Army Research Office for support provided through Grant No.\ 
DAAD-19-02-1-0051. I would also like to thank the National Center for
Supercomputing Applications at the University of Illinois at Urbana-Champaign
for access to the TeraGrid cluster through Grant No.\ PHY040024T.
\end{acknowledgments}

\end{document}